\pdfoutput=1
\documentclass[a4paper,11pt]{article}

\usepackage{lineno}
\usepackage{jheppub}
\makeatletter
\def\@fpheader{\phantom{Prepared for submission to JHEP}}
\makeatother

\usepackage[utf8]{inputenc}
\usepackage{amsmath}
\usepackage{amsthm}
\usepackage{amssymb}
\usepackage{booktabs} 
\usepackage{enumitem}   
\usepackage[english]{babel}
\usepackage{url}
\usepackage{mathtools}
\usepackage{bbold}
\usepackage{slashed}
\usepackage{multirow}
\usepackage{lipsum}
\usepackage{xcolor}
\usepackage[dvipsnames]{xcolor}
\usepackage[table]{xcolor}
\usepackage{float}
\usepackage{revsymb}
\usepackage{braket}
\usepackage{siunitx}
\usepackage{xfrac}
\usepackage{physics}
\usepackage[subpreambles=true]{standalone}
\usepackage{comment}

\allowdisplaybreaks

\newcommand{\vevv}[1]{\langle #1\rangle}

\begin{document}


\title{Thermodynamic constraints on warm inflation with the Standard Model}

\author[a,b]{Martin A. Mojahed}

\affiliation[a]{INFN Sezione di Roma, Piazzale Aldo Moro 2, I-00185 Rome, Italy}
\affiliation[b]{Università degli Studi di Roma La Sapienza, Piazzale Aldo Moro 5, I-00185 Rome, Italy}


\abstract{In warm inflation with the Standard Model (SM), an axion-like inflaton dissipates energy into the plasma through QCD sphaleron transitions. Current predictions assume a friction linear in the bias, defined as the free energy released per transition in units of the temperature. The friction, however, has only been simulated at moderate biases in pure SU(2) gauge theory. We show that, during slow roll, energy conservation, the axial anomaly and thermodynamics relate the bias to the quark chemical potentials generated by the transitions, independently of the form of the friction and of the reaction network in the plasma. A bias within the simulated range requires chemical potentials a few times the temperature, which typically lead to Higgs condensation. Conversely, small chemical potentials imply a large bias and, unless the dissipation is very weak, a temperature far above the axion decay constant, which requires a cutoff of the effective description above that temperature. We illustrate that deviations from linearity beyond the simulated range can change the inflationary dynamics qualitatively, so that reliable predictions require a real-time computation of the friction and the noise in QCD at large bias. With high-scale lepton-number violation, the quark chemical potentials also source a $B-L$ asymmetry that exceeds the observed value by many orders of magnitude already during inflation. Finally, we show that the same relation between the bias and the chemical potentials holds for derivative couplings of the inflaton to SM currents. Unless the dissipation is very weak, warm inflation driven by such couplings therefore requires either a temperature far above the decay constant or chemical potentials that typically lead to Higgs condensation.}

\maketitle


\section{Introduction}
\label{sec:intro}

Cosmic inflation~\cite{Starobinsky:1980te,Guth:1980zm,Linde:1981mu,Mukhanov:1981xt,Albrecht:1982wi,Bardeen:1983qw} has become the leading explanation for the initial conditions of the hot big bang, but its connection to particle physics remains unknown. In warm inflation~\cite{Berera:1995ie}, interactions between the inflaton and a thermal plasma generate a friction force that causes the inflaton to dissipate energy into the plasma throughout inflation. This friction allows slow roll on steeper potentials, while thermal fluctuations of the plasma enhance the scalar perturbations and thereby suppress the tensor-to-scalar ratio. The friction can also allow inflation to end smoothly in a radiation-dominated universe~\cite{Berera:2008ar,Kamali:2023lzq}. A longstanding challenge for warm inflation has been the construction of consistent microscopic realizations, since the couplings required to sustain dissipation generically spoil the flatness of the inflaton potential~\cite{Yokoyama:1998ju,Berera:2008ar}. This obstruction can be overcome in symmetry-protected constructions~\cite{Bastero-Gil:2016qru,Berghaus:2019whh}, including models with an axion-like coupling to a non-Abelian gauge field, in which the shift symmetry protects the potential and sphaleron transitions in the plasma generate the friction~\cite{McLerran:1990de,Berghaus:2019whh,Laine:2021ego}. Until recently, these realizations required new sectors beyond the Standard Model (SM). A minimal alternative was proposed in Ref.~\cite{Berghaus:2025dqi}, where the SM gluons provide the gauge field. The key observation was that cosmic expansion dilutes~\cite{Drewes:2023khq} the quark chiral asymmetry, which would otherwise suppress the friction~\cite{McLerran:1990de,Berghaus:2020ekh,Berghaus:2024zfg}. With the inflaton as the only new field and its couplings to the SM fully specified, the scenario is unusually predictive and an interesting starting point for connecting inflation with the baryon asymmetry of the Universe (BAU) or dark matter (DM)~\cite{Chen:2025try}. The proposal has since been considered in several works, see e.g.~\cite{ORamos:2025uqs,Broadberry:2025ggb,Laine:2025rll,Rogelj:2026kgr,Biondini:2026wfy,Dejrah:2026ltr,Sereewat:2026kbc,Ghoshal:2026qio}.

The rolling inflaton biases the sphaleron transitions: each transition releases a free energy, which in units of the temperature we call the bias. Current predictions of SM warm inflation (SMWI) rely on the friction at linear order in this bias. The dependence of the friction on the bias has been simulated only in pure SU(2) gauge theory, for biases up to $24$, and found to be linear up to about $6$ and to deviate mildly beyond about $10$~\cite{Moore:1996qs}. The benchmark points of Ref.~\cite{Berghaus:2025dqi}, in contrast, operate at biases of $\mathcal{O}(10^2\cdots 10^3)$, where most estimates predict a gluon instability. It is therefore important to establish whether such large biases are specific to particular solutions or inherent to the scenario, and how sensitively the predictions depend on deviations from the linear law.

Each sphaleron transition also converts right-handed quarks into left-handed ones, thereby generating quark chemical potentials that are diluted by the expansion. We show that large biases are unavoidable unless these chemical potentials exceed the temperature by a factor of a few. In the quasi-stationary state reached during slow roll, energy conservation, the axial anomaly, and the thermodynamics of the plasma alone relate the bias to the chemical potential of the right-handed up quarks, for any inflaton potential and plasma response. A bias within the simulated range then requires chemical potentials at which the Higgs doublet is typically expected to condense. Unless the dissipation is very weak, small chemical potentials require a temperature, and hence a cutoff of the effective description, far above the axion decay constant $f$. Moreover, we illustrate how modifying the response beyond the simulated range, while keeping it close to the linear law within it, can change the inflationary dynamics qualitatively: a benchmark point that is warm with the linear law can become effectively cold. Since the response beyond the simulated range is unknown, reliable predictions require a real-time computation of the friction and the noise in QCD.

The quark chemical potentials also make SMWI sensitive to the origin of neutrino masses. If the light neutrino masses arise from physics at or above the inflationary temperature, as in high-scale seesaw mechanisms~\cite{Minkowski:1977sc,Yanagida:1979as,Yanagida:1980xy,Gell-Mann:1979vob,Mohapatra:1979ia,Schechter:1980gr,Magg:1980ut,Lazarides:1980nt,Mohapatra:1980yp,Cheng:1980qt,Foot:1988aq,Ma:1998dn}, the associated lepton-number-violating interactions are also active during inflation. Through spontaneous leptogenesis~\cite{Cohen:1987vi,Cohen:1988kt,Kuckenberg:2026oax}, the quark chemical potentials then generate a $B-L$ asymmetry, which we show to exceed the observed value by several orders of magnitude already at horizon crossing.

Finally, we show that the relation between the bias and the chemical potentials is not specific to the coupling to gluons, but extends to derivative couplings of the inflaton to other currents. Examples within or near the SM are couplings to the right-handed top-quark current or, as for a majoron, to the $B-L$ current. Large chemical potentials then typically lead to Higgs condensation, whereas small ones require a temperature far above $f$ unless the dissipation is very weak.

While this work was being completed, Ref.~\cite{Ghoshal:2026qio} appeared, reaching similar conclusions about the size of the bias and the validity of the effective-operator description. Where the two analyses overlap, the results agree. Beyond the overlap, we show that a small bias requires large chemical potentials for any network of reactions in the plasma, even with a slowly relaxing gluon helicity. Since the top Yukawa interaction transfers part of the quark chirality to the Higgs doublet, we also study Higgs condensation and spontaneous leptogenesis. In addition, we examine the sensitivity of the inflationary evolution to the response and extend the relations to derivative couplings of the inflaton to SM currents.

Sec.~\ref{sec:model} introduces the model and reviews the sphaleron friction. Sec.~\ref{sec:relation} derives the relations between the bias and the quark chemical potentials, Sec.~\ref{sec:beyond} applies them to the benchmark points of Ref.~\cite{Berghaus:2025dqi}, Sec.~\ref{sec:pheno} discusses the potential generation of a $B-L$ asymmetry, and Sec.~\ref{sec:general} extends the relations to derivative couplings of the inflaton to other currents. We conclude in Sec.~\ref{sec:conclusions}. App.~\ref{app:db} derives the linear law from detailed balance, and App.~\ref{app:transport} presents the transport equations and their stationary solutions. The discussion of the friction in Sec.~\ref{sec:friction} and of the gluon instability in Sec.~\ref{sec:beyond} applies to any gauge sector heated by an axion-like inflaton, whereas for sphaleron heating the relations of Sec.~\ref{sec:relation} require a charge that only the sphaleron transitions change appreciably, such as the chirality of the light quarks of the SM. Throughout, we assume, as in Ref.~\cite{Berghaus:2025dqi}, that a thermal state is reached during inflation (see Refs.~\cite{Ferreira:2017lnd,Berghaus:2019whh,Laine:2021ego,DeRocco:2021rzv,Fujita:2025zoa,Bhattacharya:2025guc,Berghaus:2026jkw} for discussions of how such a state can arise). Whether kinetic equilibrium is then maintained is discussed in Sec.~\ref{sec:heat}.

\section{Model}
\label{sec:model}
The model of Ref.~\cite{Berghaus:2025dqi} is defined by
\begin{equation}
\label{eq:L}
\mathcal{L}=\frac12\,\partial_\mu\phi\,\partial^\mu\phi-V(\phi)-\frac{\phi}{f}\,q+\mathcal{L}_{\rm SM}\,,
\qquad q\equiv\frac{\alpha_s}{8\pi}\,G^a_{\mu\nu}\tilde G^{a\mu\nu}\,,
\end{equation}
where $G^a_{\mu\nu}$ is the gluon field strength, $\tilde G^{a\mu\nu}=\frac12\epsilon^{\mu\nu\rho\sigma}G^a_{\rho\sigma}$, $\alpha_s$ is the strong coupling and $V(\phi)$ is the inflaton potential, which in the numerical examples of Ref.~\cite{Berghaus:2025dqi} takes the form
$V=\lambda\phi^4$. The operator $q$ is the topological charge density of the gluon field. For a classical,
homogeneous inflaton in a spatially flat Friedmann universe, Eq.~\eqref{eq:L} gives the equation of
motion~\cite{McLerran:1990de,Berghaus:2025dqi}
\begin{equation}
\label{eq:eom}
\ddot\phi+3H\dot\phi+V'(\phi)=-\frac{\vevv{q}}{f}\,,
\end{equation}
where $\vevv{q}$ is the expectation value of $q$ in the state of the plasma. When the inflaton is at rest and the plasma is in full thermal equilibrium, $\vevv{q}$ vanishes, up to instanton effects that are exponentially suppressed at the temperatures of interest~\cite{Gross:1980br}. By contrast, a rolling inflaton induces a nonzero expectation value, $\vevv{q}\neq 0$. It is standard to define the friction coefficient $\Upsilon$ by $\vevv{q}\equiv f\,\Upsilon\dot\phi$, so that the right-hand side of Eq.~\eqref{eq:eom} is the force $-\Upsilon\dot\phi$, which acts as friction for $\Upsilon>0$. In
warm inflation its strength is measured by the ratio $Q\equiv\Upsilon/(3H)$, with weak ($Q<1$) and strong ($Q>1$) dissipation.\footnote{If the inflaton were a standard axion, whose potential is generated by instantons and respects the discrete shift symmetry $\phi\to\phi+2\pi f$, slow-roll inflation with strong dissipation would be excluded~\cite{Zell:2024cyz}.}

The energy which the inflaton loses through the friction is transferred to the plasma. The interaction term in Eq.~\eqref{eq:L} does not contribute to the energy-momentum tensor, so that the total energy density is the sum of $\rho_\phi=\dot\phi^2/2+V(\phi)$ and the energy density $\rho_R$ of the plasma. It follows that
\begin{equation}
\label{eq:energy}
\dot\rho_\phi+3H\dot\phi^2=-\frac{\dot\phi}{f}\,\vevv{q}\,, \qquad \dot\rho_R+3H(\rho_R+p_R)=\frac{\dot\phi}{f}\,\vevv{q}\,,
\end{equation}
with $p_R$ the pressure of the plasma. Eq.~\eqref{eq:eom} and \eqref{eq:energy} hold for any homogeneous
state of the plasma, with all details about dissipation encoded in $\vevv{q}$. 

\subsection{Friction from sphaleron heating}
\label{sec:friction}
Since $\int d^3x\,q=\dot N_{\rm CS}$, the friction is set by the mean drift of the Chern--Simons (CS) number per unit volume, which in general requires a non-perturbative real-time computation. Each unit change in $N_{\rm CS}$, corresponding to a sphaleron transition, changes the number of left-handed quarks of each flavor by $c_i=+1$ and that of right-handed quarks by $c_i=-1$~\cite{tHooft:1976rip,Kuzmin:1985mm,Rubakov:1996vz}. At weak coupling, sphaleron transitions are parametrically slower than the gauge interactions, $\alpha_s^5T$ versus $\alpha_s^2T$~\cite{Arnold:1996dy,Arnold:2002zm}. Between transitions, the plasma thus thermalizes at fixed quark numbers, with a chemical potential $\mu_i$ for each.

The rolling inflaton acts as a chemical potential $\dot\phi/f$ for $N_{\rm CS}$~\cite{Cohen:1987vi,McLerran:1990de,Kuzmin:1992up}. Each sphaleron transition transfers the energy $\dot\phi/f$ from the inflaton to the plasma, and creating the quarks costs the free energy $\mu_{\rm sph}\equiv\sum_ic_i\mu_i$. In units of the temperature, the free energy dissipated in the plasma per transition is the bias
\begin{equation}
\label{eq:bias}
A\equiv\eta-\mu_{\rm sph}/T\,,\qquad \eta\equiv\dot\phi/(fT)\,.
\end{equation}
If the dynamics is time-reversal invariant and $\dot\phi$ is approximately constant while the plasma thermalizes, detailed balance fixes the ratio of the forward and backward transition rates to $e^A$~\cite{Khlebnikov:1988sr,Moore:1996qs}. At linear order in the bias, the drift is then set by the equilibrium CS diffusion rate $\Gamma_{\rm sph}(\mu_i)$ of a plasma with the quark chemical potentials $\mu_i$,
\begin{equation}
\label{eq:q-full}
\langle q\rangle=\tfrac12\,\Gamma_{\rm sph}(\mu_i)\,A\,\big[1+O(A\mu_i/T)+O(A^2)\big]\,,
\end{equation}
see App.~\ref{app:db} for a derivation. The leading term, $\langle q\rangle=\tfrac12\,\Gamma_{\rm sph}(\mu_i)\,A$, is referred to as the linear law~\cite{McLerran:1990de,Drewes:2023khq,Berghaus:2025dqi}. In the absence of quark asymmetries, it gives the friction coefficient $\Upsilon_{\rm sph}=\Gamma_{\rm sph}/(2f^2T)$, where $\Gamma_{\rm sph}\equiv\Gamma_{\rm sph}(0)$ is known from the lattice~\cite{Bodeker:1999gx,Moore:2010jd}.

Detailed balance does not determine the response beyond linear order, which depends on how the bias modifies the dynamics of the soft gauge fields. It is known only from classical real-time simulations of hot SU(2) gauge theory without fermions~\cite{Moore:1996qs}, in which the drift is linear up to $A\simeq6$, turns up beyond $A\sim10$, and exceeds the linear law by $10\cdots25\%$ at $A=15\cdots24$, the largest bias simulated. The data turn up where the bias is estimated to become an order-one perturbation of the soft gauge dynamics, $A\sim\pi^2$~\cite{Moore:1996qs}. For SU(3), linearity was assumed rather than verified~\cite{Moore:1996qs}, and larger biases have, to our knowledge, not been simulated in a thermal plasma.

The chemical potential $\mu_{\rm sph}$ must include every charge that the transitions change and that faster processes do not erase. In the SM, these include the chiralities of the quarks whose Yukawa interactions are slow. The gluon helicity may also contribute, as each transition may change it by an amount of order $1/\alpha_s$ and it may relax slowly~\cite{Broadberry:2025ggb,Dejrah:2026ltr}. In a linearized treatment, the helicity then lowers $Q$ by a factor of order one for strong dissipation~\cite{Dejrah:2026ltr,Ghoshal:2026qio}. Our derivations do not include the helicity explicitly, but we show at the end of Sec.~\ref{sec:entropy} that it does not weaken the lower bounds on the bias and the inflaton velocity derived there.

The quark asymmetries built up by the transitions reduce the bias, and hence the friction. If the asymmetries relax to their quasi-steady values faster than $\dot\phi$ and $T$ change, the friction coefficient defined below Eq.~\eqref{eq:eom} is~\cite{Drewes:2023khq}
\begin{equation}
\label{eq:Yflip}
\Upsilon_{\rm eff}=\Upsilon_{\rm sph}\,\frac{\gamma_{\rm flip}}{\gamma_\chi+\gamma_{\rm flip}}\,,
\end{equation}
where $\gamma_\chi\equiv6N_f\Gamma_{\rm sph}/(N_cT^3)$ is the rate at which the transitions drive $\mu_{\rm sph}$ towards $\dot\phi/f$, and $\gamma_{\rm flip}$ is the rate at which Hubble dilution and chirality-flipping processes erase the asymmetries. If the asymmetries relax more slowly, $\langle q\rangle$ depends on the history of $\dot\phi$, and the asymmetries must be evolved together with the inflaton, as described in App.~\ref{app:transport}.

\paragraph{The friction coefficient of SMWI:} Ref.~\cite{Berghaus:2025dqi} evaluates Eq.~\eqref{eq:Yflip} for $N_f=5$ flavors with a common chemical potential $\mu_{jR}=-\mu_{jL}=\mu_A$, so that $\mu_{\rm sph}=-2N_f\mu_A$, and includes dilution from Hubble expansion, $\gamma_{\rm flip}=3H$. For small asymmetries this gives~\cite{Berghaus:2025dqi}
\begin{equation}
\label{eq:Yeff}
\Upsilon_{\rm eff}=
\frac{\Upsilon_{\rm sph}}{1+\dfrac{\Upsilon_{\rm sph}}{3H}\,\dfrac{N_f}{N_c}\,\dfrac{12f^2}{T^2}},
\end{equation}
which is the friction coefficient on which SMWI is built.

Eq.~\eqref{eq:Yeff} rests on two expansions, which are not on the same footing. The expansion in the chemical potentials is a simplification rather than a limitation, because at linear order in the bias the asymmetries enter only through equilibrium quantities: the thermodynamics of the quarks and the rate $\Gamma_{\rm sph}(\mu_i)$. For an ideal gas of massless quarks, $n_i=N_c\mu_i(T^2+\mu_i^2/\pi^2)/6$ per chirality, and including the cubic term merely replaces $H$ by $H\,[1+\mu_A^2/(\pi^2T^2)]$ in Eq.~\eqref{eq:Yeff}. The dependence of $\Gamma_{\rm sph}$ on the chemical potentials has, to our knowledge, not been computed on the lattice, but a weak-coupling estimate suggests that the rate drops by about $10\%$ at $|\mu_u|=T$ and by a factor of $1.8$ at $|\mu_u|=2.7\,T$.\footnote{At weak coupling and zero bias, the chemical potentials affect the soft gluon dynamics through the Debye mass~\cite{Bodeker:1998hm,Arnold:1998cy}, $m_D^2=2g^2T^2+g^2\sum_i\mu_i^2/(4\pi^2)$ for six massless flavors, with the sum over left- and right-handed components. The quoted reductions follow from inserting $m_D$ into the rate of Ref.~\cite{Moore:2010jd}, with $|\mu_i|=|\mu_u|$ for both chiralities of the five lightest flavors as in Ref.~\cite{Berghaus:2025dqi}.} The expansion in the bias, in contrast, is a genuine limitation, since $\langle q\rangle$ is not known beyond the simulated range.


\section{General relations and bounds}
\label{sec:relation}

The friction coefficient in Eq.~\eqref{eq:Yeff} is derived under two assumptions: a small bias and small chemical potentials. In this section, we show that the two assumptions exclude each other in the quasi-stationary state reached during slow roll. The argument rests on two relations, both independent of the response of the plasma. Energy conservation, the axial anomaly, and the thermodynamics of the plasma relate the inflaton velocity $\eta$ to the chemical potential of the right-handed up quarks, and entropy balance relates the bias to the same chemical potential.

\subsection{Heat and chirality}
\label{sec:heat}

By the axial anomaly, each unit increase of $N_{\rm CS}$ removes one right-handed quark of each flavor, so that the net number density $n_u$ of right-handed up quarks, summed over colors, obeys
\begin{equation}
\label{eq:nu}
\dot n_u+3Hn_u=-\langle q\rangle\,.
\end{equation}
This result neglects chirality flips induced by the up-quark Yukawa coupling, whose rate is suppressed by $y_u^2$ and far below the Hubble rate during SMWI. For the same reason, it also holds for the right-handed down and strange quarks. Eq.~\eqref{eq:nu} also neglects the hypercharge anomaly, whose effect is negligible for the benchmark points of Ref.~\cite{Berghaus:2025dqi}, as shown in App.~\ref{app:stationary}.

During slow roll, the inflaton velocity and the Hubble rate vary little over an $e$-fold, while the energy density and the quark asymmetries of the plasma relax within about one $e$-fold to the values at which production balances dilution~\cite{Berera:2008ar,Kamali:2023lzq}. In the resulting quasi-stationary state, the time derivatives in Eqs.~\eqref{eq:energy} and~\eqref{eq:nu} are negligible,
\begin{equation}
\label{eq:stationary}
3H(\rho_R+p_R)=\frac{\dot\phi}{f}\,\langle q\rangle\,,\qquad
3Hn_u=-\langle q\rangle\,.
\end{equation}
Combining the two equations gives a central relation of this paper,
\begin{equation}
\label{eq:enthalpy}
n_u=-\frac{f}{\dot\phi}\,(\rho_R+p_R)\,.
\end{equation}
Each unit increase of $N_{\rm CS}$ transfers the energy $\dot\phi/f$ to the plasma and removes one right-handed up quark, while the expansion dilutes the energy density and the asymmetry in right-handed up quarks at the rates $3H(\rho_R+p_R)$ and $3H|n_u|$. In the quasi-stationary state, production balances dilution, so that the plasma carries the enthalpy $\dot\phi/f$ per missing right-handed up quark. Eq.~\eqref{eq:enthalpy} follows from Eqs.~\eqref{eq:energy} and~\eqref{eq:nu}, which hold for any homogeneous state of the plasma, and from quasi-stationarity alone.\footnote{Without quasi-stationarity, the right-hand side of Eq.~\eqref{eq:enthalpy} acquires the factor $[1+\dot\rho_R/(3H(\rho_R+p_R))]/[1+\dot n_u/(3Hn_u)]$, which differs from one by terms of the order of the slow-roll parameters. Departure from quasi-stationarity is discussed further in Sec.~\ref{sec:numerics}.} It thus relies neither on kinetic equilibrium nor on any property of $\langle q\rangle$.

If, in addition, the chemical potentials are small and the plasma is in kinetic equilibrium, the leading-order ideal-gas expressions $\rho_R+p_R=(2\pi^2/45)\,g_*T^4$ and $n_u=N_c\mu_uT^2/6$ combined with Eq.~\eqref{eq:enthalpy} yield
\begin{equation}
\label{eq:eta}
\eta=c\,\frac{T}{|\mu_u|}\,,\qquad c\equiv\frac{4\pi^2g_*}{15N_c}\simeq93.7\,,
\end{equation}
with $\mu_u<0$ for $\eta>0$. Small chemical potentials thus require a fast inflaton, $\eta\gg c$. The value of $c$ assumes kinetic equilibrium of all SM species, $g_*=106.75$. If instead only the quarks and gluons are in kinetic equilibrium and the other species carry negligible energy, then $g_*=79$ and $c\simeq69$.

Kinetic equilibrium requires the plasma to thermalize faster than the universe expands. At leading order in kinetic theory, the shear viscosity of the quark-gluon plasma~\cite{Arnold:2003zc} implies a thermalization rate $\Gamma_{\rm th}\simeq(N_c\alpha_s)^2T\simeq0.006\,T$~\cite{York:2008rr}, whereas Refs.~\cite{Laine:2021ego,Fujita:2025zoa} assume a rate ten times larger, chosen large on purpose~\cite{Laine:2021ego}. With the kinetic-theory estimate, the strongly dissipative benchmark points of Tab.~\ref{tab:bdz} ($Q>1$) have $\Gamma_{\rm th}/H\simeq3\cdots 50$, and the weakly dissipative ones ($\lambda=10^{-17}$, $10^{-16}$ and $10^{-15}$) have $\Gamma_{\rm th}/H\simeq1$, $0.3$ and $0.1$. With the larger rate, every ratio grows by a factor of ten. For the weakly dissipative points, kinetic equilibrium during inflation is therefore not assured.\footnote{Ref.~\cite{Ghoshal:2026qio} reaches a similar conclusion by requiring the soft magnetic modes that mediate the transitions to relax faster than the expansion.} Without kinetic equilibrium, the temperature and the chemical potentials are not defined and Eq.~\eqref{eq:Yeff} does not apply, whereas Eq.~\eqref{eq:enthalpy} still holds.

\subsection{Entropy and the bias}
\label{sec:entropy}
The balance of heat and chirality in Sec.~\ref{sec:heat} shows that small chemical potentials require a fast inflaton, $\eta\gg c$. The friction coefficient in Eq.~\eqref{eq:Yeff}, however, is derived to linear order in the bias $A=\eta-\mu_{\rm sph}/T$, which the quark asymmetries reduce below $\eta$. A fast inflaton therefore does not by itself imply a large bias. In this section, we determine $A$ from the entropy balance of the plasma. Each transition dissipates the free energy $TA$ in the plasma (Sec.~\ref{sec:friction}) and thus produces the entropy $A$, while the expansion dilutes the entropy. In the quasi-stationary state, entropy production balances dilution, so that the bias equals the entropy of the plasma per missing right-handed up quark, up to small corrections. The resulting relation holds for chemical potentials of any size and bounds the bias from below in terms of the chemical potential of the right-handed up quarks.

For a plasma in kinetic equilibrium, the Euler and Gibbs relations give
\begin{equation}
\label{eq:entropyrate}
T(\dot s+3Hs)=\dot\rho_R+3H(\rho_R+p_R)-\sum_i\mu_i(\dot n_i+3Hn_i)\,.
\end{equation}
The number densities $n_i$ also change through reactions other than strong sphaleron transitions, including electroweak sphaleron transitions, Yukawa interactions, and possibly BSM interactions.  Following Ref.~\cite{Domcke:2020kcp}, we label these reactions by $\alpha$ and characterize each by the numbers $n_i^\alpha$ of particles of species $i$ that a single reaction creates ($n_i^\alpha>0$) or destroys ($n_i^\alpha<0$). We further define the net reaction rate per unit volume, $J_\alpha$, and the affinity $\mathcal A_\alpha\equiv-\frac1T\sum_in_i^\alpha\mu_i$, which is the free energy released per reaction in units of $T$. The strong sphaleron transitions have the net rate $\langle q\rangle$, the numbers $n_i^{\rm sph}=c_i$ and the affinity $A$. The number densities thus obey
\begin{equation}
\label{eq:transport}
\dot n_i+3Hn_i=n_i^{\rm sph}\langle q\rangle+\sum_\alpha n_i^\alpha J_\alpha\,,
\end{equation}
and inserting Eqs.~\eqref{eq:energy} and~\eqref{eq:transport} into Eq.~\eqref{eq:entropyrate} gives the entropy balance
\begin{equation}
\label{eq:entropybalance}
\dot s+3Hs=A\,\langle q\rangle+\sum_\alpha\mathcal A_\alpha J_\alpha\,.
\end{equation}
Each strong sphaleron transition produces the entropy $A$ and each reaction of type $\alpha$ the entropy $\mathcal A_\alpha$. Every term on the right-hand side is non-negative, since every reaction proceeds on average in the direction in which it releases free energy.

Let $\delta$ denote the fraction of the entropy production in Eq.~\eqref{eq:entropybalance} that comes from reactions other than the strong sphaleron transitions. In the quasi-stationary state, where $\langle q\rangle=3H|n_u|$, Eq.~\eqref{eq:entropybalance} then reduces to
\begin{equation}
\label{eq:Abias}
A=\frac{s}{|n_u|}\,(1-\delta)\,,
\end{equation}
for any response $\langle q\rangle$. The correction $\delta$ is small, because the other reactions produce entropy only at second order in the chemical potentials, whereas the dilution term $3Hs$ is nonzero for $\mu_i\to0$ and proportional to $g_*$. For the stationary solution of App.~\ref{app:stationary}, $\delta\lesssim5\mu_u^2/(\pi^2g_*T^2)\simeq5\times10^{-3}\,\mu_u^2/T^2$, cf. Eq.~\eqref{eq:deltabound}.

Our results for $\eta$ and $A$ in Eqs.~\eqref{eq:enthalpy} and~\eqref{eq:Abias}, respectively, are related through the Euler relation,
\begin{equation}
\label{eq:etaA}
\eta=\frac{\rho_R+p_R}{T|n_u|}=\frac{A}{1-\delta}+\sum_i\left(\frac{\mu_i}{T}\right)\left(\frac{n_i}{|n_u|}\right)\,.
\end{equation}
Since $\sum_i\mu_in_i\geq0$, the bias bounds the inflaton velocity from below, $\eta\geq A/(1-\delta)$. If the chemical potentials are small and the quark numbers are the only slowly relaxing charges that the transitions change, Eq.~\eqref{eq:etaA} implies that $\eta$ and $A$ agree up to corrections of relative order $\mu_u^2/T^2$. For the reaction network of App.~\ref{app:transport}, $\eta-A=\mu_{\rm sph}/T$ lies between roughly $10\,|\mu_u|/T$ and $12\,|\mu_u|/T$, so that the relative difference is about $0.1\,\mu_u^2/T^2$. If one assumes the linear law, $\Upsilon_{\rm eff}=\langle q\rangle/(f\dot\phi)=\Upsilon_{\rm sph}A/\eta$, then the quark asymmetries reduce the friction by the same relative amount, which stays below $10\%$ at the benchmark points of Tab.~\ref{tab:bdz}. Hubble dilution thus prevents the quarks from suppressing the friction, as found in Refs.~\cite{Drewes:2023khq,Berghaus:2025dqi}.

Finally, for small chemical potentials, $s\simeq(2\pi^2/45)\,g_*T^3$, and Eq.~\eqref{eq:Abias} becomes
\begin{equation}
\label{eq:main}
A\simeq c\,\frac{T}{|\mu_u|}\,,
\end{equation}
with the same $c$ as in Eq.~\eqref{eq:eta}. Small chemical potentials thus require a large bias, $A\gg c$. Eqs.~\eqref{eq:eta} and~\eqref{eq:main} also follow from the stationary limit of the equations of Refs.~\cite{Berghaus:2025dqi,Drewes:2023khq}, which assume the linear law and small chemical potentials, whereas Eqs.~\eqref{eq:enthalpy} and~\eqref{eq:Abias} require neither assumption.

\paragraph{Chemical potentials of arbitrary size:}
Eq.~\eqref{eq:main} suggests that a small bias requires large chemical potentials, where its validity breaks down. To extend the analysis to this regime, we use Eqs.~\eqref{eq:Abias} and~\eqref{eq:etaA}, which hold for arbitrary chemical potentials, to derive general lower bounds on $A$ and $\eta$. The bounds require the full chemical-potential dependence of $|n_u|$, $s$, and $\sum_i\mu_in_i$: for massless quarks, the density of right-handed up quarks is $|n_u|=(N_c|\mu_u|T^2/6)\,[1+\mu_u^2/(\pi^2T^2)]$, and each quark chirality with $|\mu_i|=|\mu_u|$ adds $N_c\mu_u^2T/6$ to the entropy density and $|\mu_u||n_u|$ to $\sum_i\mu_in_i$.

In general, the generated asymmetries may be redistributed among the species in the plasma through a network of reactions. However, the numbers of the $N_{f'}\geq3$ lightest right-handed quarks change only through the strong sphaleron transitions, since their Yukawa interactions are negligible, cf. Sec.~\ref{sec:heat}. Their chemical potentials are therefore also equal to $\mu_u$. The chemical potentials of all other species contribute non-negatively to $s$ and to $\sum_i\mu_in_i$. Neglecting these contributions in Eqs.~\eqref{eq:Abias} and~\eqref{eq:etaA} yields the lower bounds
\begin{equation}
\label{eq:bound}
\frac{A}{1-\delta}\geq\frac{cT/|\mu_u|+N_{f'}|\mu_u|/T}{1+\mu_u^2/(\pi^2T^2)}\,,\qquad
\eta\geq\frac{A}{1-\delta}+N_{f'}\,\frac{|\mu_u|}{T}\,.
\end{equation}
The right-hand side of the first inequality decreases monotonically with $|\mu_u|/T$ and, for $N_{f'}=3$, equals $88$, $24$ and $6$ at $|\mu_u|/T=1$, $2.75$ and $6.8$, where $24$ is the largest simulated bias and $6$ the largest bias with a verified linear response, cf. Sec.~\ref{sec:friction}. With $\delta\lesssim5\times10^{-3}\mu_u^2/T^2$, the bias thus exceeds $87$ for $|\mu_u|\leq T$. Provided that $\delta$ remains small beyond the validity of this leading-order estimate, $A\leq24$ requires $|\mu_u|\gtrsim2.7\,T$. The two inequalities also bound the inflaton velocity from below, $\eta>24$, with the minimum of the bound near $|\mu_u|\simeq5\,T$.

Finally, if the anomaly is the only process that changes particle numbers, then $\delta=0$, and both chiralities of the $N_f$ massless flavors carry $|\mu_u|$. All other chemical potentials vanish, and Eq.~\eqref{eq:bound} holds with equality and $N_{f'}$ replaced by $2N_f$. The resulting relation for the bias reproduces Eq.~(19) of Ref.~\cite{Ghoshal:2026qio}.

\paragraph{Impact of slowly relaxing gauge helicity:} As discussed in Sec.~\ref{sec:friction}, each sphaleron transition may also change the gluon helicity by an amount $\nu_h\sim1/\alpha_s$, and the helicity may relax more slowly than the universe expands~\cite{Broadberry:2025ggb,Dejrah:2026ltr}. A helicity that can be described by a density $n_h$ and a chemical potential $\mu_h$~\cite{Dejrah:2026ltr} enters our relations as an additional charge, which adds $\nu_h\mu_h$ to $\mu_{\rm sph}$ and non-negative contributions to $s$ and $\sum_i\mu_in_i$, while the perturbative processes that change the helicity contribute about one per cent to $\delta$ for the parameters of Ref.~\cite{Dejrah:2026ltr}. Eqs.~\eqref{eq:Abias}, \eqref{eq:etaA} and~\eqref{eq:bound} therefore remain valid, and Eq.~\eqref{eq:enthalpy} holds irrespective of the helicity dynamics. Eqs.~\eqref{eq:eta} and~\eqref{eq:main}, as well as the approximate equality $\eta\simeq A$, additionally require the helicity contributions to be negligible, which is not guaranteed even for $|\mu_u|\ll T$, since $n_h\simeq\nu_h|n_u|\gg|n_u|$ in the stationary state. The helicity can therefore make the inflaton substantially faster at a given $\mu_u$, but cannot lower the bias or the inflaton velocity below the bounds~\eqref{eq:bound}.

\subsection{Numerical example solutions}
\label{sec:numerics}
The relations of Secs.~\ref{sec:heat} and~\ref{sec:entropy} hold for any response $\langle q\rangle$, and Eqs.~\eqref{eq:Abias}, \eqref{eq:etaA} and~\eqref{eq:bound} even for chemical potentials of arbitrary size. All of these relations, however, assume a quasi-stationary state, which must break down towards the end of inflation. Moreover, they relate the bias to the chemical potentials at each moment, but do not determine how the inflaton and the temperature evolve, which depends on the response. To quantify the departures from quasi-stationarity and to illustrate the dependence on the response, we solve the transport equations of App.~\ref{app:transport} for the homogeneous evolution of the inflaton, the temperature and the chemical potentials, using the parameters $(\lambda,f)$ of the seven benchmark points of Ref.~\cite{Berghaus:2025dqi}. Unlike the relations above, these equations require a choice of response, for which we use the linear law unless stated otherwise. Like Eq.~\eqref{eq:Yeff}, they are also linearized in the chemical potentials. Neither approximation is justified throughout inflation, so that the solutions illustrate the description underlying the model's predictions, rather than its actual evolution.

\begin{figure}[t]
\centering
\includegraphics[width=0.8\textwidth]{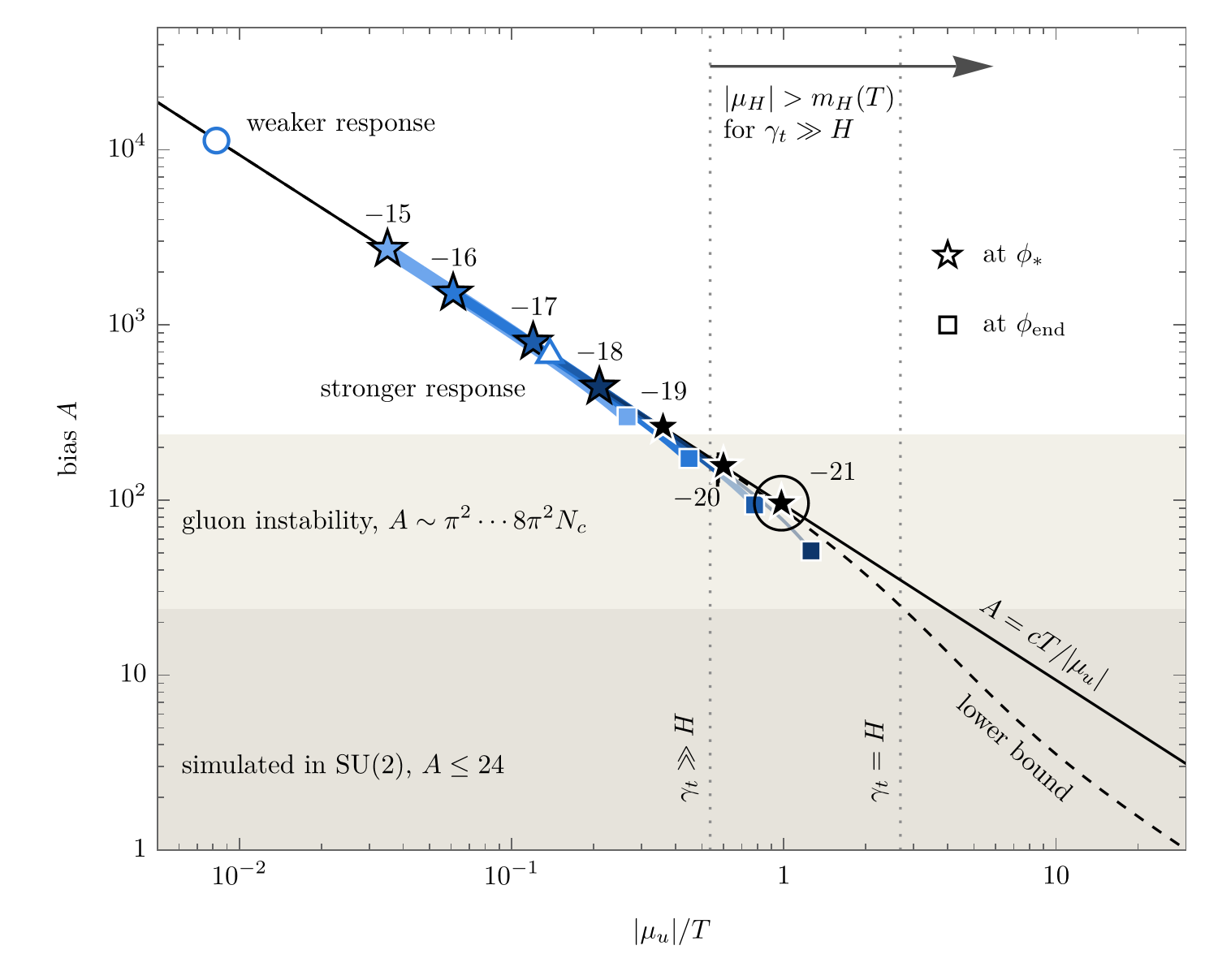}
\caption{Bias $A$ of the strong sphaleron transitions against $|\mu_u|/T$. Solid line: Eq.~\eqref{eq:main}, satisfied by every stationary state with small chemical potentials, for any $f$, inflaton potential and response. Dashed line: the lower bound~\eqref{eq:bound} on $A/(1-\delta)$ in any stationary state. Shading: biases simulated in pure SU(2) gauge theory, with a linear response up to $A\simeq6$~\cite{Moore:1996qs}. Horizontal band: estimates of the bias above which the response becomes non-linear or the gluon field unstable, from $A\simeq\pi^2$~\cite{Moore:1996qs} to $8\pi^2N_c$ (Sec.~\ref{sec:beyond}). Dotted lines: estimated $|\mu_u|$ above which $|\mu_H|>m_H(T)$ and the Higgs doublet condenses in a stationary state, for $\gamma_t\gg H$ (left) and $\gamma_t=H$ (right). Stars: the benchmark points of Tab.~\ref{tab:bdz} at horizon crossing, labeled by $\log_{10}\lambda$; the circled star has $|\mu_H|>m_H(T)$. Colored lines: the solutions of Sec.~\ref{sec:numerics} for $\lambda=10^{-15}\cdots10^{-18}$ from horizon crossing to the end of inflation (squares), drawn lighter where $|\mu_H|>m_H(T)$. The squares lie below the dashed line because the solutions are no longer stationary and the equations of App.~\ref{app:transport} use the leading-order thermodynamics, which is inaccurate for $|\mu_u|\gtrsim T$. Open symbols: the solution for $\lambda=10^{-16}$ at $\phi_*$ with the weaker (circle) and stronger (triangle) response of Eq.~\eqref{eq:responses}.}
\label{fig:bias}
\end{figure}

Fig.~\ref{fig:bias} shows the solutions for $\lambda=10^{-15}\cdots10^{-18}$, from horizon crossing of the CMB pivot scale, at the field value $\phi_*$ of Ref.~\cite{Berghaus:2025dqi}, to the end of inflation (squares), when the expansion stops accelerating. For $\lambda\leq10^{-19}$, for which the Higgs doublet condenses well before the end of inflation, we show only the benchmark points at horizon crossing (stars). The solutions follow the solid line, Eq.~\eqref{eq:main}: between $\phi_*$ and the end of inflation, $|\mu_u|/T$ rises from $0.035\cdots0.21$ to $0.27\cdots1.26$, while $A$ falls from $2650\cdots440$ to $299\cdots51$, remaining above the simulated range throughout. Since the transport equations use the leading-order thermodynamics, a stationary state satisfies Eq.~\eqref{eq:eta} exactly, and the deviation of $\eta|\mu_u|/(cT)$ from one measures the time derivatives neglected in Eqs.~\eqref{eq:stationary}. The ratio is $0.997\cdots0.998$ at $\phi_*$ for all seven points, remains within $1\%$ of one until about $15$ $e$-folds and within $5\%$ until about two $e$-folds before the end, and drops to $0.85\cdots0.86$ at the end of inflation.

For $\lambda\le10^{-17}$, $|\mu_H|$ exceeds the thermal Higgs mass $m_H(T)$ before the end of inflation, at $|\mu_u|\simeq0.57\,T$. This is close to the threshold for a top-Yukawa rate $\gamma_t\gg H$ (left dotted line), beyond which the Higgs doublet condenses and the equations cease to apply, as discussed in Sec.~\ref{sec:beyond}. As long as the Higgs doublet remains uncondensed, the bias remains above $A\simeq145$.

Although the relations derived in Sec.~\ref{sec:relation} do not depend on $\langle q\rangle$, the evolution of the inflaton and the temperature does. The solutions above involve biases well beyond the range in which the linear law has been tested. We therefore examine the sensitivity of the evolution to the response at large bias. For $\lambda=10^{-16}$, we repeat the calculation using weaker and stronger responses, both of which remain close to the linear law over the simulated range,
\begin{align}
\label{eq:responses}
    \langle q\rangle=\tfrac12\Gamma_{\rm sph}A\,(1+|A|/100)^{-1/2}, \qquad \langle q\rangle=\tfrac12\Gamma_{\rm sph}A\,(1+|A|/240).
\end{align} 
Both responses are odd in $A$, as required by CP invariance and explained in App.~\ref{app:db}. They deviate from the linear law by less than $3\%$ for $A\le6$ and by $10\%$ at $A=24$, with the weaker response lying below and the stronger one above the linear prediction. By comparison, the drift observed in SU(2) exceeds the linear law by up to $25\%$ at $A\sim 15\cdots 24$~\cite{Moore:1996qs}.\footnote{Whether this excess persists in SU(3) and at larger bias is unknown. Responses that, like the SU(2) data, exceed the linear law over the simulated range, but fall below it at larger bias, can give results similar to those of the weaker response considered here. For example, $\langle q\rangle=\tfrac12\Gamma_{\rm sph}A\,[1+(A/50)^2]/[1+(|A|/57)^{5/2}]$, which exceeds the linear law by $10\%$ at $A=24$ and falls below it for $A>96$, gives $Q=3.3\times10^{-5}$ and $T/H=8.5$ at $\phi_*$.} At $A=1.5\times10^3$, corresponding to the bias of the linear-law solution at $\phi_*$, the two responses differ from the linear law by factors of $4$ and $7$, respectively. For the same $\lambda$ and $f$, the number of $e$-folds between $\phi_*$ and the end of inflation becomes $50.4$ and $96.0$, instead of $60.3$. At $\phi_*$, the weaker response gives $Q=3.6\times10^{-5}$ and $T/H=8.7$, instead of $Q=0.098$ and $T/H=60$, so that dissipation becomes negligible for the inflaton evolution and, with $Q\ll1$, too weak to thermalize the inflaton fluctuations.\footnote{Ref.~\cite{Berghaus:2025dqi} finds $Q=0.0889$ and $T/H=58.9$ at $\phi_*$. The values differ mainly because the sphaleron rate used there is a few per cent smaller than ours, cf. App.~\ref{app:transport}, and the scaling discussed below amplifies the effect on $Q$. With the rate of Ref.~\cite{Berghaus:2025dqi}, our equations reproduce these values to within a few per cent. Ref.~\cite{Berghaus:2025dqi} also finds fewer $e$-folds after $\phi_*$, $57.5$ instead of $60.3$, because its rate is lower and because it computes both the $e$-folds and the end of inflation in the slow-roll approximation.} The stronger response gives $Q=0.83$ and $T/H=79$. At $60.3$ $e$-folds before the end of inflation, where the linear-law solution passes through $\phi_*$, the two responses give $Q=2.3\times10^{-5}$ and $0.86$, respectively. These large changes arise because, at fixed $\phi$ and up to the running of $\alpha_s$ and departures from stationarity, a response that differs from the linear law by a factor $G$ changes the dissipation ratio according to $Q(1+Q)^6\propto G^4$.\footnote{This scaling follows from the slow-roll equation $3H(1+Q)\dot\phi=-V'$, the first of Eqs.~\eqref{eq:stationary}, and $Q\propto G\,T^3$, which holds for $\Gamma_{\rm sph}\propto T^4$.} Nevertheless, both states lie on the solid line in Fig.~\ref{fig:bias} (open symbols), at $A\simeq1.1\times10^4$ and $680$, instead of $1.5\times10^3$. Changing the response thus moves a state along the line $A\simeq cT/|\mu_u|$, but not off it. The position on this line, and hence the predictions for given $\lambda$ and $f$, is determined by the response at large bias, which is not known.

Fitting $\lambda$ and $f$ to observations does not, in general, remove the dependence on the response. A constant rescaling of the linear response, with the noise rescaled accordingly, could be absorbed into $f$, provided the quark asymmetries remain too small to affect the dynamics. During inflation, however, the bias falls by an order of magnitude, 
so that the ratio of the response to the linear law need not remain constant over the relevant range, and two parameters cannot absorb an arbitrary dependence on the bias. Moreover, beyond linear order, the friction entering the perturbations is determined by $\partial\langle q\rangle/\partial\dot\phi$, rather than $\langle q\rangle/\dot\phi$, and the noise is no longer fixed by the friction~\cite{Ghoshal:2026qio}.

\section{Implications for sphaleron heating}
\label{sec:beyond}
Current predictions for SMWI rely on the friction coefficient~\eqref{eq:Yeff}, which assumes the linear law. As shown in Sec.~\ref{sec:relation}, however, the benchmark points of Ref.~\cite{Berghaus:2025dqi} operate at biases well above the simulated range, where the evolution depends sensitively on the unknown response. Since a bias within the simulated range would require $|\mu_u|\gtrsim2.7\,T$, SMWI operates either at large chemical potentials or at a large bias, and this section discusses the consequences of each regime. Large chemical potentials typically lead to Higgs condensation, whereas small chemical potentials imply a large bias, at which the gluon field can become unstable. Unless the dissipation is very weak, small chemical potentials also require a temperature far above $f$. Except for the Higgs doublet, the same points were discussed in Ref.~\cite{Ghoshal:2026qio}, and our results agree where the analyses overlap.

Tab.~\ref{tab:bdz} lists the benchmark points at horizon crossing of the CMB pivot scale, with $\eta$, $|\mu_u|$ and $A$ obtained from the stationary relations alone. The bias ranges from $97$ to $2.7\times10^3$. With the linear-response law, the first of Eqs.~\eqref{eq:stationary} yields
\begin{align}
    \eta A=\frac{8H\rho_R}{\Gamma_{\rm sph}T}=\frac{4\pi^2g_*}{15}\,\frac{H/T}{\kappa_{SS}\alpha_s^5},
\end{align}
i.e.\ $A\simeq10^4\,(H/T)^{1/2}$ for $A\simeq\eta$, in agreement with Eq.~(16) of Ref.~\cite{Ghoshal:2026qio}. Hence, a bias in the simulated range would require $T/H\gtrsim10^5$. A hidden gauge sector with a larger coupling would operate at a smaller bias, since $A\propto\alpha^{-5/2}$ at fixed $T/H$. Even for $N_c=3$ and $\alpha=0.1$, however, Ref.~\cite{Ghoshal:2026qio} estimates $A\approx25\cdots100$. Independently of the response, Eq.~\eqref{eq:bound} allows a bias within the simulated range, $A\le24$, only for $|\mu_u|\gtrsim2.7\,T$.

\begin{table}[t]
\centering
\begin{tabular}{ccccccccc}
\hline
$\lambda$ & $Q$ & $T/H$ & $T/f$ & $\eta$ & $|\mu_u|/T$ & $A$ & $|\mu_H|/T$ & $k_{\rm inst}/T$\\
\hline
$10^{-21}$ & $14.9$   & $7820$ & $60$ & $106$  & $0.88$  & $97$   & $0.57$   & $0.42$\\
$10^{-20}$ & $9.22$   & $3551$ & $72$ & $163$  & $0.57$  & $157$  & $0.36$   & $0.66$\\
$10^{-19}$ & $4.78$   & $1474$ & $85$ & $266$  & $0.35$  & $263$  & $0.20$   & $1.1$\\
$10^{-18}$ & $2.19$   & $582$  & $96$ & $445$  & $0.21$  & $443$  & $0.095$  & $1.8$\\
$10^{-17}$ & $0.661$  & $202$  & $95$ & $801$  & $0.12$  & $800$  & $0.033$  & $3.1$\\
$10^{-16}$ & $0.0889$ & $58.9$ & $66$ & $1525$ & $0.061$ & $1525$ & $0.0072$ & $5.9$\\
$10^{-15}$ & $0.0122$ & $18.8$ & $44$ & $2710$ & $0.035$ & $2710$ & $0.0015$ & $10$\\
\hline
\end{tabular}
\caption{The benchmark points of Ref.~\cite{Berghaus:2025dqi} at horizon crossing of the CMB pivot scale. The values of $\lambda$, $Q$ and $T/H$ are taken from Table~I of the published version of Ref.~\cite{Berghaus:2025dqi}, and $T/f$ from the values of $T$ and $f$ given there. The other columns follow from the stationary relations at leading order in the chemical potentials, without the Weinberg operator: $\eta$ from $Q$ and $T/f$ via the first of Eqs.~\eqref{eq:stationary}, $|\mu_u|$ from Eq.~\eqref{eq:eta}, and $A$ and $|\mu_H|$ from App.~\ref{app:stationary} with the top-Yukawa rate of Tab.~\ref{tab:reactions}. The Higgs doublet condenses if $|\mu_H|$ exceeds the thermal Higgs mass, $m_H(T)\simeq0.36\,T$. For $k_{\rm inst}$, we use the values of $\alpha_s$ of Ref.~\cite{Berghaus:2025dqi}.}
\label{tab:bdz}
\end{table}

\paragraph{Higgs condensation:} The top-Yukawa interaction transfers part of the quark chirality to the Higgs doublet, $|\mu_H|=r\,|\mu_u|$, where $r$ grows with $\gamma_t/H$ from $0$ to $2/3$, as detailed in App.~\ref{app:stationary}. In a stationary state, $|\mu_H|$ thus exceeds the thermal Higgs mass $m_H(T)$ for $|\mu_u|>m_H(T)/r$, and the excess asymmetry condenses, breaking the electroweak symmetry in the plasma~\cite{Linde:1976kh,Benson:1991nj,Bajc:1997ky}.\footnote{For $g_2=0.55$, $g_Y=0.42$ and $y_t=0.49$, $m_H(T)\simeq(3g_2^2/16+g_Y^2/16+y_t^2/4)^{1/2}\,T\simeq0.36\,T$, so condensation requires $|\mu_u|>0.54\,T$. At leading order in the chemical potentials, a bias within the simulated range is compatible with an uncondensed Higgs doublet only for $r\lesssim0.13$, corresponding to $\gamma_t\lesssim H$ (Fig.~\ref{fig:bias}) and thus $T/H\lesssim70$. In this regime, kinetic equilibrium is itself not guaranteed, cf. the discussion at the end of Sec.~\ref{sec:heat}.} A condensate would not affect the strong sphaleron transitions, and Eq.~\eqref{eq:main} would hold with a smaller $c$. However, for the central values of $m_t$ and $\alpha_s$, the Higgs quartic coupling is negative at the benchmark temperatures of Ref.~\cite{Berghaus:2025dqi}, $T\simeq10^{13}\cdots4\times10^{14}$\,GeV, and would be positive for $m_t$ smaller by about $1$ to $2$\,GeV~\cite{Buttazzo:2013uya,Hiller:2024zjp}. The sign of the quartic coupling at these temperatures is thus sensitive to the theoretical uncertainty in relating the measured top mass to the top Yukawa coupling~\cite{Alekhin:2012py,Hoang:2020iah}. With a negative quartic coupling, only the thermal mass holds the Higgs field at the origin~\cite{Espinosa:2007qp}, and $\mu_H$ lowers the squared mass to $m_H^2(T)-\mu_H^2$. The chemical-potential-induced reduction of the Higgs mass is reminiscent of a different mechanism, where top quarks produced through a derivative coupling to the inflaton lower the squared Higgs mass and break electroweak symmetry in cold inflation~\cite{Hook:2019vcn}. There, the Higgs vacuum expectation value is limited by the top mass it generates. Whether the condensate induced by $\mu_H$ is similarly limited has, to our knowledge, not been studied. Pending a dedicated analysis, we regard $|\mu_H|<m_H(T)$ as a consistency condition of the scenario. The condition is violated at horizon crossing for $\lambda=10^{-21}$ and marginal for $\lambda=10^{-20}$ (Tab.~\ref{tab:bdz}), and fails before the end of inflation for all $\lambda\le10^{-17}$.

\paragraph{A tachyonic instability:} Independently of the quarks, there is a concrete reason to doubt the linear law at large bias. Acting as a chemical potential for $N_{\rm CS}$, the bias makes one circular polarization of the gluon modes with $k<k_{\rm inst}\equiv\alpha_sT|A|/(2\pi)$ grow exponentially, as a rolling axion does for Abelian gauge fields~\cite{Anber:2009ua}. In a non-Abelian plasma, magnetic screening is expected to remove the instability as long as $k_{\rm inst}$ lies below the magnetic screening mass $m_M$~\cite{Fujita:2025zoa}, which requires $A\lesssim4\pi^2\simeq40$ for the lattice value $m_M\simeq0.5\,g^2T$ in SU(2)~\cite{Heller:1997nqa} and SU(3)~\cite{Nakamura:2003pu}, and $A\lesssim8\pi^2N_c\simeq237$ for the estimate $m_M\simeq g^2N_cT$ of Ref.~\cite{Fujita:2025zoa}. A model of the dispersion relation of the soft modes places the threshold at $A\approx80\cdots90$~\cite{Ghoshal:2026qio}. At horizon crossing, all benchmark points of Tab.~\ref{tab:bdz} lie above $80\cdots90$, and the five above $8\pi^2N_c$ have $k_{\rm inst}>T$, so that the instability reaches even hard gluons. For $\lambda=10^{-16}$ and $10^{-15}$, the fastest-growing modes, at $k=k_{\rm inst}/2\simeq3\,T$ and $5\,T$, grow at a rate of about $k$, far above the damping rate of hard gluons. It is then unclear whether a near-thermal state with a well-defined sphaleron rate exists. Even where such a state exists, the effect of the instability on the friction remains open. Ref.~\cite{Fujita:2025zoa} argues that the warm-inflation attractor survives an unscreened instability, because the radiation it produces stays below that supplied by the friction, but assumes the linear law. Real-time simulations find that a decaying chirality imbalance destabilizes non-Abelian gauge fields in a plasma~\cite{Akamatsu:2015kau,Schlichting:2022fjc} and drives a drift of $N_{\rm CS}$ roughly proportional to the initial imbalance~\cite{Akamatsu:2015kau}. For an axion condensate coupled to SU(2) gauge fields without a thermal bath, the gauge self-interactions appear to moderate the instability, and the axion is damped at a rate quadratic in its coupling, as for the sphaleron friction~\cite{Berghaus:2026jkw}. Neither setup maintains a bias in a thermal plasma, as the inflaton does in warm inflation, so that the stationary drift at biases of order $10^2\cdots 10^3$ has not been determined.

\paragraph{Scale separation:} The stationary relations also tie the quark chemical potential to the scale $f$. With $\langle q\rangle=3HQf\dot\phi$, Eqs.~\eqref{eq:stationary} give
\begin{align}
    \label{eq:nuf}
    |n_u|=f\,[Q(\rho_R+p_R)]^{1/2},
\end{align}
for any response of the plasma. For an ideal gas in kinetic equilibrium, the relation becomes
\begin{equation}
\label{eq:muf}
|\mu_u|\simeq(6cQ/N_c)^{1/2}f\simeq14\,Q^{1/2}f,
\end{equation}
up to corrections of relative order $\mu_u^2/(\pi^2T^2)$. Keeping $|\mu_u|$ below $T$ thus requires $T\gtrsim14\,Q^{1/2}f$, and hence $T>f$ unless $Q\lesssim5\times10^{-3}$. The benchmark points all have larger $Q$, with $T/f\simeq44\cdots96$, cf. Tab.~\ref{tab:bdz}. Ref.~\cite{Berghaus:2025dqi} lists $f$ and $T$ at these points and shows that $T$ lies at least an order of magnitude below the cutoff of its effective description, $8\pi f/\alpha_s$. As pointed out in Ref.~\cite{Ghoshal:2026qio}, however, $8\pi f/\alpha_s$ is an upper estimate of the cutoff. Adopting it assumes that the particles generating the inflaton--gluon coupling are at least as heavy. For $T\gg f$, a UV completion is difficult to construct even if these particles need only be heavier than $T$. In a minimal renormalizable Peccei--Quinn (PQ) completion~\cite{Peccei:1977hh}, for instance, the inflaton is the Goldstone boson of the PQ symmetry, broken at the scale $v_{\rm PQ}=\mathcal N f$ with $\mathcal N$ the anomaly coefficient. Keeping the PQ symmetry broken in the plasma then requires $\mathcal N\gtrsim T/f\sim10^2$ and, for unit PQ charges, as many color-triplet fermions, which drive the strong coupling to a Landau pole below the Planck scale~\cite{Ghoshal:2026qio}.

\section{Large chemical potentials and spontaneous leptogenesis}
\label{sec:pheno}
The relations derived in Sec.~\ref{sec:relation} imply that sphaleron heating is necessarily accompanied by non-vanishing quark chemical potentials. During slow roll, Eq.~\eqref{eq:muf} fixes their size in terms of $Q$ and $f$, for any response of the plasma. The top Yukawa interaction transfers part of this chiral asymmetry to the Higgs doublet. In the presence of $B-L$-violating interactions, these chemical potentials can source a $B-L$ asymmetry through spontaneous leptogenesis driven by the inflaton~\cite{Cohen:1987vi,Cohen:1988kt,Kuckenberg:2026oax}; see also Refs.~\cite{Brandenberger:2003kc,Bastero-Gil:2011clw,Basak:2021cgk} for baryogenesis during warm inflation.

This observation is particularly relevant if light neutrino masses originate from a high-scale seesaw mechanism~\cite{Minkowski:1977sc,Yanagida:1979as,Yanagida:1980xy,Gell-Mann:1979vob,Mohapatra:1979ia,Schechter:1980gr,Magg:1980ut,Lazarides:1980nt,Mohapatra:1980yp,Cheng:1980qt,Foot:1988aq,Ma:1998dn}. For definiteness, we consider the simplest regime in which all BSM states mediating lepton-number violation (LNV) are much heavier than the highest temperature reached during the evolution. Their thermal production is then negligible, and LNV is described by the dimension-five Weinberg operator~\cite{Weinberg:1979sa}, through which the Higgs chemical potential sources a $B-L$ asymmetry. We find that the asymmetry exceeds the observed value by several orders of magnitude already at horizon crossing. States with masses around the inflationary temperature would make LNV faster and, since the stationary asymmetry grows with the LNV rate, only increase it. The asymmetry grows further towards the end of inflation and shortly after, where our description is no longer under control. Our slow-roll estimate therefore does not determine the final asymmetry, which also depends on entropy production and possible washout after inflation.

During slow roll, the lepton asymmetry approaches a quasi-stationary value at which its production by the Weinberg operator balances its dilution. Applying the leading-order results of App.~\ref{app:stationary}, we obtain
\begin{equation}
\label{eq:BLs}
Y_{B-L}\equiv \frac{n_{B-L}}{s}=\frac{15K}{\pi^2g_*}\,\frac{\mu_H}{T}\,,
\end{equation}
where $K\in(0,3/2)$, defined in App.~\ref{app:stationary}, increases with the LNV reaction rates relative to the Hubble rate. The Higgs chemical potential is related to the right-handed up-quark chemical potential by $\mu_H=r\mu_u$, where $r\in(0,2/3)$ increases with $\gamma_t/H$, cf.~Eq.~\eqref{eq:muH}, and $\mu_u$ is fixed by Eq.~\eqref{eq:muf}. Thus, at fixed $f$, the asymmetry depends on the plasma response only through the background quantities $Q$, $T$ and $H$, which the Weinberg operator does not affect appreciably. We can therefore estimate the asymmetry from the background values at horizon crossing listed in Tab.~\ref{tab:bdz}, but compute $\mu_H$ with the Weinberg operator, which lowers $|\mu_H|$ by up to $15\%$. We normalize our results to the value corresponding to the observed BAU,
\begin{equation}
\label{eq:YBLobs}
Y_{B-L}^{\text{obs}}=c_{\text{sph}}^{-1}Y_B^{\text{obs}},
\end{equation}
where $Y_B^{\text{obs}}\simeq 8.7\times10^{-11}$~\cite{Planck:2018vyg} and we take the electroweak sphaleron conversion factor~\cite{Harvey:1990qw,Laine:1999wv} to be $c_{\text{sph}}=0.3328$~\cite{Benso:2026csa}. We vary the lightest neutrino mass over $m_1\in[0,0.03]\,\mathrm{eV}$, the range allowed for normal ordering by $\sum_i m_i<0.12\,\mathrm{eV}$~\cite{Planck:2018vyg}. Tab.~\ref{tab:B-L} shows that the asymmetry at horizon crossing exceeds the observed value by factors of $2.7\times 10^3\cdots 7.7\times 10^5$ for all benchmark points, with the smallest values attained at $m_1=0$. Upon solving the transport equations in App.~\ref{app:transport} numerically, we have verified that the asymmetry grows as a function of $e$-folds. This is because dissipation becomes stronger as inflation proceeds, which raises $\mu_u/T$ by Eq.~\eqref{eq:muf}, while the expansion rate decreases faster than the rates of the top-Yukawa interaction and the Weinberg operator. Towards the end of inflation, however, these transport equations cease to be reliable: the chemical potentials approach the temperature, and for most benchmark points the Higgs doublet condenses.

\begin{table}[t]
\centering
\renewcommand{\arraystretch}{1.4}
\setlength{\tabcolsep}{8pt}
\begin{tabular}{clcccccc}
\toprule
 & $\lambda$ & $10^{-15}$ & $10^{-16}$ & $10^{-17}$ & $10^{-18}$ & $10^{-19}$ & $10^{-20}$\\
\midrule
\multirow{2}{*}{$\dfrac{1}{10^4}\,\dfrac{Y_{B-L}}{Y_{B-L}^{\rm obs}}$} & $m_1=0$ & $0.27$ & $2.7$ & $18$ & $35$ & $38$ & $33$\\
 & $m_1=0.03$\,eV & $0.54$ & $5.2$ & $33$ & $67$ & $77$ & $66$\\
\bottomrule
\end{tabular}
\caption{Quasi-stationary $B-L$ asymmetry at horizon crossing, given by Eq.~\eqref{eq:BLs} and normalized to $10^4$ times the observed baryon asymmetry, Eq.~\eqref{eq:YBLobs}. Results are shown for the benchmark points in Tab.~\ref{tab:bdz}, considering the smallest and largest values of the lightest neutrino mass allowed by $\sum_i m_i<0.12\,\mathrm{eV}$~\cite{Planck:2018vyg} for normal ordering. The point $\lambda=10^{-21}$ is omitted because the Higgs doublet has already condensed at horizon crossing.}
\label{tab:B-L}
\end{table}

In summary, sphaleron heating necessarily generates quark asymmetries that exceed the observed baryon asymmetry by many orders of magnitude. Since reactions such as the top Yukawa interaction transfer part of these asymmetries to other species, $B-L$-violating interactions that are not much slower than the Hubble expansion generically source a $B-L$ asymmetry above the observed value. For a high-scale seesaw, the Weinberg operator alone yields an asymmetry exceeding the observed value by several orders of magnitude already at horizon crossing. The final asymmetry is set near and after the end of inflation, where our description is not under control. Avoiding overproduction would require, for instance, substantial entropy production or a washout that remains efficient after the quark asymmetries have been erased.

\section{Generalization to derivative couplings}
\label{sec:general}
The challenges identified above for SMWI, including a bias beyond the simulated range, a large $T/f$, and Higgs condensation, raise the question of whether alternative inflaton couplings to the SM can avoid them. To address this question, we first observe that the relations of Sec.~\ref{sec:relation} are not specific to sphaleron heating, but rest on two properties: each unit change of a charge (the number of right-handed up quarks in SMWI) transfers the energy $\dot\phi/f$ from the inflaton to the plasma, and the inflaton transfers energy in no other way. Both properties are preserved if we replace the inflaton--SM coupling in Eq.~\eqref{eq:L} with a derivative coupling $(\partial_\mu\phi/f)\,j_M^\mu$ to the current $j_M^\mu$ of a charge $M$. After integration by parts, the inflaton then couples to every operator that violates $M$, including Yukawa interactions and, for an anomalous charge, terms such as $G\tilde G$, $W\Tilde{W}$, and $Y\tilde Y$. The coupling to $Y\tilde Y$ can amplify helical hypermagnetic fields directly, whereas in SMWI such fields arise only through the chiral plasma instability, which is slow at horizon crossing (App.~\ref{app:stationary}). We neglect such fields here. For a homogeneous inflaton, the equation of motion and energy conservation give
\begin{equation}
\label{eq:eomM}
\ddot\phi+3H\dot\phi+V'(\phi)=-\frac{\dot n_M+3Hn_M}{f}\,,\qquad
\dot\rho_R+3H(\rho_R+p_R)=\frac{\dot\phi}{f}\,(\dot n_M+3Hn_M)\,.
\end{equation}
It follows that Eqs.~\eqref{eq:enthalpy}, \eqref{eq:Abias} and~\eqref{eq:nuf} remain valid for any plasma response, with $n_u$ replaced by $-n_M$, $A$ by the mean bias per unit change of $M$, and $\delta$ by the entropy fraction from reactions that conserve $M$. Eqs.~\eqref{eq:Abias} and~\eqref{eq:nuf} then fix the asymmetry per unit entropy, $|n_M|/s\simeq1/A\simeq0.15\,Q^{1/2}f/T$.

Within the SM, coupling the inflaton to the current of a charge that only sphaleron transitions change appreciably, such as the number of right-handed up quarks, leads back to sphaleron heating. The right-handed top-quark current offers a different possibility, since $n_{t_R}$ changes mainly through the top Yukawa interaction, whose rate is several orders of magnitude faster than that of strong sphalerons. Another possibility is identifying the inflaton with the majoron, the Goldstone boson of $B-L$~\cite{Chikashige:1980ui}, which couples to the $B-L$ current. Unlike the charges above, $B-L$ is free of anomalies and changes only through interactions beyond the SM, namely those that generate the neutrino masses.

If $M$ is the number of right-handed top quarks, the faster top-Yukawa interaction reduces the bias at fixed $T/H$ and transfers the asymmetry to the Higgs doublet, $|n_H|\simeq|n_M|$. Avoiding Higgs condensation then requires $A\gtrsim2\times 10^2$ and $T/f\gtrsim30\,Q^{1/2}$. A coupling to the right-handed top-quark current is thus subject to essentially the same constraint as SMWI.

If $M=B-L$, the minimal majoron model gives $f=2v_{B-L}$, with $v_{B-L}$ the $B-L$-breaking scale. The effective description then applies only while the $B-L$ symmetry remains broken, which requires $T\lesssim f/2$. If, as in Sec.~\ref{sec:pheno}, $B-L$ violation is induced by the Weinberg operator, the asymmetry resides in the lepton doublets. For lepton chemical potentials below the temperature, Eq.~\eqref{eq:nuf} gives $T\gtrsim7\,Q^{1/2}f$. Compatibility with $T\lesssim f/2$ then requires $Q\lesssim5\times10^{-3}$. For larger $Q$, a standard majoron is pushed to large chemical potentials and, since the Weinberg operator also changes the Higgs number, typically to Higgs condensation. Moreover, the $B-L$ asymmetry generated during inflation, $|n_{B-L}|/s\simeq1/A$, exceeds the BAU by many orders of magnitude.

Thus, barring very weak dissipation, warm inflation driven by a derivative coupling to a current whose charge is violated by interactions within or near the SM faces the same alternative as SMWI. Small chemical potentials entail a large bias and $T\gg f$, so that the cutoff of the effective description, such as the symmetry-breaking scale of a Goldstone boson, must lie well above $f$. A bias of order one instead requires chemical potentials well above the temperature, which can induce a Higgs condensate if interactions such as the top Yukawa coupling transfer part of the asymmetry to the Higgs doublet. Whether such a condensate is compatible with warm inflation remains to be studied, cf. Sec.~\ref{sec:beyond}. The relations of Sec.~\ref{sec:relation} do not apply to charge-preserving couplings, such as the mass modulation in the Warm Little Inflaton~\cite{Bastero-Gil:2016qru,Levy:2020zfo}.

\section{Conclusions}
\label{sec:conclusions}
Current predictions of SMWI rely on a plasma response $\langle q\rangle$ that is linear in the bias, although this response has been simulated only in pure SU(2) gauge theory and only for biases up to $24$~\cite{Moore:1996qs}. We have shown that, in the quasi-stationary state reached during slow roll, energy conservation, the axial anomaly and the thermodynamics of the plasma fix the bias as the entropy of the plasma per missing right-handed up quark, up to small corrections. This relation holds for any inflaton potential, $f$, response and network of reactions in the plasma, and explicit solutions confirm it until about two $e$-folds before the end of inflation. Small quark chemical potentials thus necessarily come with a large bias. In the SM plasma, keeping the bias within the simulated range requires $|\mu_u|\gtrsim2.7\,T$, even when a slowly relaxing gluon helicity is included, and at such chemical potentials the Higgs doublet is typically expected to condense. 

The benchmark points of Ref.~\cite{Berghaus:2025dqi}, with $|\mu_u|<T$ at horizon crossing, operate at biases of $\mathcal{O}(10^2\cdots10^3)$. At such biases, the predictions depend sensitively on the response beyond linear order. As illustrated in Sec.~\ref{sec:numerics}, responses that remain close to the linear law within the simulated range, but deviate from it at larger bias, can change the inflationary dynamics qualitatively, for instance turning a warm benchmark point into an effectively cold one. The biases of the benchmark points also exceed most estimates of the threshold above which the gluon field becomes unstable, and the corresponding values $T/f\simeq44\cdots96$ require a cutoff of the effective description far above $f$. For the weakly dissipative points, even kinetic equilibrium is not assured. With the linear law, moreover, the Higgs chemical potential exceeds the thermal Higgs mass at horizon crossing for $\lambda=10^{-21}$ and before the end of inflation for all $\lambda\le10^{-17}$. The biases are large because the sphaleron transitions are rare: to compensate for the dilution of the plasma by the expansion, each transition must release a large free energy. Since the sphaleron rate scales as $\alpha^5T^4$, a hidden gauge sector with a larger coupling would operate at a smaller bias, and its predictions would rely less on the response beyond the simulated range. In the SM, the strong coupling is fixed, so that reliable predictions require a real-time computation in SU(3), including the quarks, of both the drift of the Chern--Simons number and its fluctuations at biases of $\mathcal{O}(10^2\cdots10^3)$.

Quark chemical potentials are a robust consequence of QCD-sphaleron heating. Since the inflaton-induced quark asymmetries exceed the BAU by many orders of magnitude, $B-L$-violating interactions that are not much slower than the Hubble expansion generically source an excessive $B-L$ asymmetry. We illustrated this point in the context of a seesaw scenario in which the heavy states have masses well above the inflationary temperature, so that LNV is mediated by the Weinberg operator with a rate that is fixed by the light neutrino masses. In this case, the resulting $B-L$ asymmetry already exceeds the observed value by three to six orders of magnitude at horizon crossing and grows further towards the end of inflation. Heavy seesaw states with masses comparable to the temperature would make LNV faster and only increase the stationary asymmetry. SMWI with high-scale LNV therefore generically overproduces the asymmetry, unless it is diluted by substantial entropy production or erased by a washout that remains efficient after the quark asymmetries have relaxed.

Finally, the relations of Sec.~\ref{sec:relation} are not specific to axions coupled to $G\tilde G$, but extend to derivative couplings of the inflaton to other currents, for any response of the plasma, as shown in Sec.~\ref{sec:general}. The asymmetry per unit entropy is then the inverse of the bias, up to small corrections. A small bias thus requires large chemical potentials, at which the Higgs doublet may condense if reactions convert part of the asymmetry into a Higgs asymmetry. Small chemical potentials, in turn, require $T\gg f$ unless the dissipation is very weak, and hence a cutoff of the effective description far above $f$. Faster reactions, such as the top Yukawa interaction, lower the bias at a given $T/H$, but only by raising the chemical potentials. Couplings that change no charge, as in the Warm Little Inflaton, are not subject to these relations.

\paragraph{Outlook:} Beyond the response at large bias, several questions about SMWI remain open. The first is the fate of the Higgs doublet. If the Higgs quartic coupling is negative at the relevant temperatures, as for the central values of $m_t$ and $\alpha_s$, the consistency of the scenario depends on whether a condensate induced by $\mu_H$ remains bounded or grows into the unstable region of the potential. As discussed in Secs.~\ref{sec:beyond} and~\ref{sec:general}, any attempt to realize axion-driven warm inflation through sphaleron or derivative couplings to the SM, while avoiding the regime $T\gg f$, must confront this issue, which motivates a dedicated study of the condensate dynamics. A second question is the post-inflationary evolution, in particular the contribution of the inflaton condensate to DM and to the effective number of relativistic species, $N_{\rm eff}$. For the quartic potential of Ref.~\cite{Berghaus:2025dqi}, the residual condensate redshifts as radiation and contributes to $N_{\rm eff}$. With an additional mass term, the condensate would redshift as matter earlier and could account for a larger fraction of the DM than found there, provided the inflaton is light enough to be long-lived, which may require tuning because the quartic term breaks the shift symmetry. Both contributions depend on the fraction of the condensate energy transferred to the plasma, and hence on the response at large bias and the dynamics at the end of inflation, which provides another motivation for a real-time computation in QCD. \\

\noindent\textbf{Acknowledgments:} The author would like to thank Mathias Becker, Kai Schmitz and Sascha Weber for discussions, Kai Schmitz, Alfredo Stanzione and Sebastian Zell for comments on the draft, and Kim Berghaus and Marco Drewes for correspondence. The author used Claude (Anthropic) to improve the wording of the manuscript, to assist with coding, and to cross-check derivations and numerical results. The author takes full responsibility for the content.

\appendix

\section{The linear law from detailed balance}
\label{app:db}
The purpose of this appendix is to derive the linear law, Eq.~\eqref{eq:q-full}, using detailed balance and CP invariance. On time scales longer than the thermalization at fixed $N_{\rm CS}$, we describe the CS number by a Markovian random walk, in which $N_{\rm CS}$ advances and retreats by one unit at the rates $\Gamma_\pm$ per unit volume, so that $\langle q\rangle=\Gamma_+-\Gamma_-$. At weak coupling, $N_{\rm CS}$ changes through the overdamped dynamics of the soft gauge fields~\cite{Arnold:1996dy,Bodeker:1998hm}, and the unit steps are a coarse-grained description of this diffusion. Including the term $-(\dot\phi/f)N_{\rm CS}$, each advance changes the free energy of the plasma by $\mu_{\rm sph}$ and the energy of the inflaton by $-\dot\phi/f$, so that the total free energy changes by $\mu_{\rm sph}-\dot\phi/f=-TA$. If the microscopic dynamics is time-reversal invariant\footnote{The dynamics is time-reversal invariant once $\bar\theta$ is absorbed into $\phi$ and CKM-induced CP violation is neglected.} and $\dot\phi$ is constant, detailed balance then gives $\Gamma_+/\Gamma_-=e^{A}$~\cite{Khlebnikov:1988sr,Moore:1996qs}.\footnote{Detailed balance requires the plasma to be thermal at each fixed $N_{\rm CS}$, but not $N_{\rm CS}$ itself to be in equilibrium, which would mean $A=0$.} With $S(A,\mu_i)\equiv\Gamma_++\Gamma_-$, it follows that
\begin{equation}
\label{eq:Stanh}
\langle q\rangle=S(A,\mu_i)\tanh(A/2)\,,
\end{equation}
so that the dynamics enters only through $S$. At $A=0$, the walk is unbiased, and $S$ is the equilibrium CS diffusion rate of a plasma with the asymmetries $\mu_i$,
\begin{equation}
\label{eq:Gammasph0}
\Gamma_{\rm sph}(\mu_i)\equiv\lim_{t\to\infty}\lim_{V\to\infty}\frac{\langle(N_{\rm CS}(t)-N_{\rm CS}(0))^2\rangle}{Vt}\bigg|_{A=0}=S(0,\mu_i)\,.
\end{equation}
Since $N_{\rm CS}$, the quark numbers and $\dot\phi$ are CP-odd, CP maps a plasma with $(A,\mu_i)$ onto one with $(-A,-\mu_i)$, and an advance onto a retreat. For CP-invariant dynamics, $S(A,\mu_i)=S(-A,-\mu_i)$, so that $\Gamma_{\rm sph}(\mu_i)$ is even in the asymmetries and, if $S$ is smooth at zero bias, $S=\Gamma_{\rm sph}(\mu_i)\,[1+O(A\mu_i/T)+O(A^2)]$. Inserted in Eq.~\eqref{eq:Stanh}, this gives Eq.~\eqref{eq:q-full}. Without asymmetries, the linear law reads $\langle q\rangle=\Gamma_{\rm sph}\dot\phi/(2fT)$, as required by the fluctuation--dissipation theorem~\cite{Kubo:1957mj,Laine:2016hma}.

Beyond linear order, the response is set by the dependence of $S$ on the bias, which detailed balance does not determine. For example, if $N_{\rm CS}$ changed through thermally activated hops over a barrier that is large compared with $T$, the bias would lower the barrier for an advance and raise it for a retreat by $TA/2$ without changing the prefactor, so that $S=\Gamma_{\rm sph}\cosh(A/2)$ and $\langle q\rangle=\Gamma_{\rm sph}\sinh(A/2)$. This relation exceeds the linear law by more than a factor of three at $A=6$, and is excluded by the simulations of Ref.~\cite{Moore:1996qs}. By contrast, for a freely diffusing variable subject to friction, the bias acts as a constant force. As long as the bias does not modify the friction or the noise, the drift is then strictly linear, $\langle q\rangle=\Gamma_{\rm sph}A/2$, which corresponds to $S=\Gamma_{\rm sph}(A/2)\coth(A/2)$~\cite{Moore:1996qs}.

\section{Transport equations}
\label{app:transport}
This appendix lists the transport equations solved numerically in Secs.~\ref{sec:numerics} and~\ref{sec:pheno}, with the LNV rates set to zero in Sec.~\ref{sec:numerics}. We also derive stationary solutions, which are used in Secs.~\ref{sec:relation},~\ref{sec:beyond} and~\ref{sec:pheno}. The transport equations are those of Refs.~\cite{Domcke:2020kcp,Kuckenberg:2026oax}, extended to a plasma that is heated by the inflaton and expressed with the number of $e$-folds, $N=\ln (a/a_*)$, as the time variable. 

\subsection{Network and equations}

\paragraph{Species and reactions:} The plasma is described by its temperature and by the chemical potentials $\mu_i$ of the following species with $g_i$ internal DoF: 
\begin{itemize}
    \item The three heaviest right-handed quarks $t$, $c$ and $b$ ($g_i=3$).
    \item The three lightest right-handed quarks $u$, $d$ and $s$ ($g_i=3$ each), which obey the same evolution equation and are represented by a single chemical potential $\mu_u$. 
    \item The quark doublets $Q_1$, $Q_2$, $Q_3$ ($g_i=6$). 
    \item The lepton doublets $L_1$, $L_2$, $L_3$ ($g_i=2$) in the basis in which the neutrino mass matrix is diagonal.
    \item The Higgs doublet $H$, for which $g_i=4$ in our conventions, because each of its two complex components contributes twice as much to the number density as a fermionic degree of freedom.
\end{itemize}
In reaction formulas and subscripts $H$ denotes the Higgs doublet; everywhere else $H$ is the Hubble rate. The Yukawa interactions of the charged leptons, whose rates remain below the Hubble rate while the Weinberg operator is active,
are omitted, so that the right-handed charged leptons carry no asymmetry. In terms of these chemical potentials,
$\mu_{\rm sph}=2(\mu_{Q_1}+\mu_{Q_2}+\mu_{Q_3})-3\mu_u-\mu_c-\mu_b-\mu_t$, where we have neglected contributions from helical gluons.

Tab.~\ref{tab:reactions} lists the reactions, which are characterized by the integers $n_i^\alpha$ introduced in Sec.~\ref{sec:entropy}, and by the affinity $\mathcal A_\alpha$. Their net rates are linear in the affinities,
\begin{equation}
\label{eq:Jalpha}
J_\alpha=\frac{T^3}{6}\,\gamma_\alpha\,\mathcal A_\alpha\,.
\end{equation}
For the strong sphaleron transitions the net rate is $\langle q\rangle$ and the affinity is the bias $A=\eta-\mu_{\rm sph}/T$, so that Eq.~\eqref{eq:Jalpha} with $\gamma_{\rm sph}=3\Gamma_{\rm sph}/T^3$ is the linear law of Sec.~\ref{sec:friction}. We use $\Gamma_{\rm sph}=\kappa_{SS}\alpha_s^5T^4$, where $\alpha_s$ is evaluated at $\mu=\pi T$ with one-loop running for six quark flavors from $\alpha_s(M_Z)=0.1184$, and $\kappa_{SS}=270\cdots274$ follows from the general-$N_c$ formula of Ref.~\cite{Moore:2010jd} evaluated with the same $\alpha_s$. Ref.~\cite{Berghaus:2025dqi} instead uses $\Gamma_{\rm sph}=(N_c\alpha_s)^5T^4$ with $\alpha_s$ evaluated at $\mu=T$, which is a few per cent smaller at the benchmark temperatures. All other couplings are kept at the values given in Tab.~\ref{tab:reactions}.
The rate coefficients of the Yukawa interactions follow the parametrization of Refs.~\cite{Garbrecht:2014kda,Domcke:2020kcp}, and that of the weak sphaleron transitions the formula of Ref.~\cite{Moore:2010jd}, both with the numerical values used in Ref.~\cite{Kuckenberg:2026oax}, where the rate coefficient of the Weinberg operator is derived.

\begin{table}[t]
\centering
\small
\begin{tabular}{@{}l@{\hspace{1.2em}}l@{\hspace{1.2em}}l@{\hspace{1.2em}}l@{}}
\toprule
reaction $\alpha$ & elementary process & affinity $\mathcal A_\alpha$ & rate coefficient $\gamma_\alpha$\\
\midrule
strong sphaleron & $q_R\to q_L$ for all six flavors & $A$ & $3\,\Gamma_{\rm sph}/T^3$\\
top Yukawa & $Q_3+H\to t$ & $(\mu_{Q_3}+\mu_H-\mu_t)/T$ & $6\times10^{-2}\,y_t^2\,T$\\
charm Yukawa & $Q_2+H\to c$ & $(\mu_{Q_2}+\mu_H-\mu_c)/T$ & $6\times10^{-2}\,y_c^2\,T$\\
bottom Yukawa & $Q_3\to b+H$ & $(\mu_{Q_3}-\mu_H-\mu_b)/T$ & $6\times10^{-2}\,y_b^2\,T$\\
weak sphaleron & $\sum_i(3\,Q_i+L_i)\to0$ & $\sum_i(3\mu_{Q_i}+\mu_{L_i})/T$ & $3\,\kappa_{\rm ws}\,\alpha_2^5\,T$\\
Weinberg operator & $L_i+L_i+H+H\to0$ & $2\,(\mu_{L_i}+\mu_H)/T$ & $9\,m_i^2\,T^3/(2\pi^5v^4)$\\
\bottomrule
\end{tabular}
\caption{Reactions included in the transport equations, with the rate coefficients
defined in Eq.~\eqref{eq:Jalpha}. The integers $n_i^\alpha$ can be read off from the
elementary process: species on the left-hand side are destroyed, those on the right-hand
side are created. A strong sphaleron transition destroys one right-handed quark of each
flavor and creates one left-handed quark of each flavor, i.e.\ two quarks in each
doublet $Q_i$. A weak sphaleron transition destroys three quarks and one lepton of each
generation. The Weinberg operator, for each $i=1,2,3$, induces $L_iL_i\to\bar H\bar H$
and $L_iH\to\bar L_i\bar H$, both of which lower the numbers of $L_i$ and of $H$ by two.
We use $y_t=0.49$, $y_c=1.8\times10^{-3}$, $y_b=6.8\times10^{-3}$, $\kappa_{\rm ws}\simeq24$
and $\alpha_2=g_2^2/4\pi$ with $g_2=0.55$ as in Ref.~\cite{Kuckenberg:2026oax}, the
neutrino masses $m_i$ for normal ordering, with $m_1=0.01$ eV unless stated otherwise,
$m_2^2-m_1^2=7.49\times10^{-5}~{\rm eV}^2$
and $m_3^2-m_1^2=2.513\times10^{-3}~{\rm eV}^2$~\cite{Esteban:2024eli}, and
$v=174$~GeV.}
\label{tab:reactions}
\end{table}

\paragraph{Evolution equations:} Let primes on $\phi$, $\rho_R$ and $H$ denote derivatives with respect to $N$, so that $\dot\phi=H\phi'$ and $\eta=H\phi'/(fT)$. Eqs.~\eqref{eq:eom} and~\eqref{eq:energy} can then be rewritten as
\begin{equation}
\label{eq:eomN}
\phi''+\Big(3+\frac{H'}{H}\Big)\phi'+\frac{V'(\phi)}{H^2}=-\frac{\langle q\rangle}{fH^2}\,,
\qquad
\rho_R'+4\rho_R=\frac{\phi'}{f}\,\langle q\rangle\,,
\end{equation}
where we have used $p_R=\rho_R/3$. The Friedmann equations give
\begin{align}
    H^2=\frac{(V+\rho_R)}{3M_P^2}\left(1-\frac{\phi'^2}{6M_P^2}\right)^{-1}, \qquad \frac{H'}{H}=-\frac{\phi'^2}{2M_P^2}-\frac{2\rho_R}{3M_P^2H^2},
\end{align}
with $M_P$ the reduced Planck mass. Using the leading-order relations for $|\mu_i|\ll T$, $n_i=g_i\mu_iT^2/6$ and $\rho_R=\pi^2g_*T^4/30$, we rewrite Eq.~\eqref{eq:transport} in terms of chemical potentials
\begin{equation}
\label{eq:transportN}
\Big(\frac{d}{dN}+\frac34\,\frac{d\ln\rho_R}{dN}+3\Big)\frac{\mu_i}{T}=\frac{6\,\mathcal C_i}{g_iHT^3}\,,
\end{equation}
where the collision terms
$\mathcal C_i=n_i^{\rm sph}\langle q\rangle+\sum_\alpha n_i^\alpha J_\alpha$ are
\begin{align}
\mathcal C_t&=-\langle q\rangle+J_t\,,&
\mathcal C_{Q_1}&=2\langle q\rangle-3J_{\rm ws}\,,\nonumber\\
\mathcal C_c&=-\langle q\rangle+J_c\,,&
\mathcal C_{Q_2}&=2\langle q\rangle-3J_{\rm ws}-J_c\,,\nonumber\\
\mathcal C_b&=-\langle q\rangle+J_b\,,&
\mathcal C_{Q_3}&=2\langle q\rangle-3J_{\rm ws}-J_t-J_b\,,\nonumber\\
\mathcal C_u&=-\langle q\rangle\,,&
\mathcal C_{L_i}&=-J_{\rm ws}-2J_{W,i}\,,\nonumber\\
&&\mathcal C_H&=-J_t-J_c+J_b-2{\textstyle\sum_i}J_{W,i}\,.
\label{eq:network}
\end{align}
Here, $\langle q\rangle=\frac{T^3}{6}\,\gamma_{\rm sph}A$ and the $J_\alpha$ follow from Eq.~\eqref{eq:Jalpha} and Tab.~\ref{tab:reactions}. The term $\frac34\,d\ln\rho_R/dN+3=3\,d\ln(aT)/dN$ is the rate at which the comoving entropy of the plasma grows and describes the dilution of $\mu_i/T\propto n_i/s$ by entropy production. It vanishes if the plasma is not heated, at constant $g_*$, so that $T\propto a^{-1}$, but is close to $3$ during warm inflation, where $\rho_R$ is quasi-stationary. Baryon and lepton number are changed by the weak sphalerons, which violate $B+L$, and by the Weinberg operator, which violates $B-L$.

\paragraph{Accuracy:} Apart from the linear law for the strong sphaleron transitions, with $\Gamma_{\rm sph}(\mu_i)$ replaced by $\Gamma_{\rm sph}$ (Sec.~\ref{sec:friction}), the equations above are linearized in the chemical potentials. The relations $n_i=g_i\mu_iT^2/6$ and $\rho_R=\pi^2g_*T^4/30$ neglect corrections of relative order $\mu_i^2/T^2$, and the net rates~\eqref{eq:Jalpha} are linear in the affinities, whereas for Boltzmann statistics the net rate of each process is proportional to the difference between the products of $n_i/n_i^{\rm eq}=e^{\mu_i/T}$ over its initial and its final states. All of these approximations rely on $|\mu_i|\ll T$. This condition is not satisfied near and after the end of inflation, where $|\mu_u|$ grows to order $T$ (Sec.~\ref{sec:numerics}), and for the most strongly dissipative benchmark points it is questionable already at horizon crossing (Tab.~\ref{tab:bdz}); there the solutions are estimates. Moreover, the equations assume a thermal distribution for every species, which for the Higgs doublet requires $|\mu_H|<m_H(T)$ (Sec.~\ref{sec:beyond}).

\subsection{Stationary solution}
\label{app:stationary}
In the quasi-stationary state deep inside the slow-roll regime, $\mu_i/T$ and $\rho_R$ are approximately constant, so that Eq.~\eqref{eq:transportN} reduces to $3Hn_i\simeq\mathcal C_i$. These are linear algebraic equations for the chemical potentials. Expressed in terms of $\mu_u$, their solution does not depend on the response $\langle q\rangle$, which is fixed by the equation for $i=u$, $\langle q\rangle\simeq3H|n_u|$. The equation for $i=u$ and the stationary energy balance together imply Eq.~\eqref{eq:eta}, and they do so exactly, because the equations use the leading-order relations for $n_i$ and $\rho_R$ which Eq.~\eqref{eq:eta} assumes. In the following, we neglect $J_{\rm ws}$, $J_c$ and $J_b$, whose rate coefficients are much smaller than the Hubble rate at horizon crossing for all benchmark points of Tab.~\ref{tab:bdz}.

Solving the linear algebraic equations, we find that all right-handed quarks other than the top carry the chemical potential $\mu_u$, that the doublets $Q_1$ and $Q_2$ carry $-\mu_u$, and that
\begin{equation}
\label{eq:mut}
\mu_t=\mu_u-\tfrac43(1+K)\,\mu_H\,,\qquad \mu_{Q_3}=-\mu_u+\tfrac23(1+K)\,\mu_H\,,\qquad
\mu_{L_i}=-\frac{2w_i}{3+2w_i}\,\mu_H\,,
\end{equation}
with $w_i\equiv\gamma_{W,i}/H$ and $K\equiv\sum_iw_i/(3+2w_i)<3/2$. The Weinberg operator thus induces chemical potentials for the lepton doublets. The top Yukawa interaction transfers part of the chirality of the quarks to the Higgs doublet, whose chemical potential is
\begin{equation}
\label{eq:muH}
\mu_H=r\,\mu_u\,,\qquad r=\frac{2\gamma_t}{12(1+K)H+(3+2K)\gamma_t}\,,
\end{equation}
where the ratio $r$ grows with $\gamma_t/H$ from $0$ to at most $2/3$. For the bias, the stationary solution yields
\begin{equation}
\eta-A=\frac{\mu_{\rm sph}}{T}=\Big[12-\tfrac83(1+K)\,r\Big]\frac{|\mu_u|}{T}\,,
\end{equation}
where the factor in brackets lies between $9.8$ and $12$.

All reactions other than the strong sphaleron transitions produce entropy only at second order in the chemical potentials. Evaluated on the algebraic solution, the top Yukawa interaction and the Weinberg operator contribute
\begin{equation}
\label{eq:delta}
\delta=\frac{45}{\pi^2g_*}\Big[\frac{4(1+K)^2H}{\gamma_t}+\sum_i\frac{w_i}{(3+2w_i)^2}\Big]\frac{\mu_H^2}{T^2}\,,
\end{equation}
where the contribution of the top Yukawa interaction dominates at the benchmark points. For $K\ll1$, Eqs.~\eqref{eq:muH} and~\eqref{eq:delta} give
\begin{equation}
\label{eq:deltabound}
\delta\simeq\frac{45}{\pi^2g_*}\,r\Big(\frac23-r\Big)\frac{\mu_u^2}{T^2}\leq\frac{5}{\pi^2g_*}\,\frac{\mu_u^2}{T^2}\,,
\end{equation}
with the maximum at $\gamma_t\simeq4H$, where $r=1/3$. The physical reason is that a fast top Yukawa interaction keeps its affinity small, while a slow one acts rarely. The reactions neglected above contribute much less to $\delta$.

The stationary solution also justifies neglecting in Eq.~\eqref{eq:nu} the hypercharge anomaly, which changes fermion numbers only in the presence of helical hypermagnetic fields. Although the inflaton couples exclusively to gluons and therefore does not directly source hypermagnetic fields, the chiral plasma instability can amplify hypermagnetic modes of one helicity with wave numbers $k<\xi$, where~\cite{Joyce:1997uy}
\begin{align}
    \xi=\frac{\alpha_Y}{\pi}\,\abs{\sum_i\varepsilon_ig_iY_i^2\mu_i}.
\end{align}
Here, $g_i$ and $Y_i$ are the multiplicity and hypercharge of species $i$, and $\varepsilon_i=+1$ ($-1$) for right- (left-)handed fermions. The stationary solution gives $\xi\leq5.5\,\alpha_Y|\mu_u|/\pi\simeq0.025\,|\mu_u|$ for $g_Y=0.42$. In a collisionless plasma, Landau damping sets the growth rate to $4k^2(\xi-k)/(\pi m_{D_Y}^2)$~\cite{Akamatsu:2013pjd}, whose maximum value is $16\xi^3/(27\pi m_{D_Y}^2)$, where $m_{D_Y}^2=11g_Y^2T^2/6$ is the hypercharge Debye mass. At horizon crossing, the maximum is largest for $\lambda=10^{-21}$, where it is at most $5\%$ of the Hubble rate for $g_Y=0.42$. Accounting for quark collisions can increase the rate, but only by a factor of order one.

Finally, without the weak sphalerons baryon number is conserved and remains zero, so that the $B-L$ asymmetry is carried by the lepton doublets alone, $\mu_{B-L}/T\equiv6n_{B-L}/T^3=4K\mu_H/T$.


\addcontentsline{toc}{section}{References}

\small

\bibliographystyle{JHEP}
\bibliography{arxiv1}

@article{Berera:1995ie,
    author = "Berera, Arjun",
    title = "{Warm inflation}",
    eprint = "astro-ph/9509049",
    archivePrefix = "arXiv",
    primaryClass = "astro-ph",
    reportNumber = "PSU-TH-159",
    doi = "10.1103/PhysRevLett.75.3218",
    journal = "Phys. Rev. Lett.",
    volume = "75",
    pages = "3218--3221",
    year = "1995"
}

@article{Berera:2008ar,
    author = "Berera, Arjun and Moss, Ian G. and Ramos, Rudnei O.",
    title = "{Warm Inflation and its Microphysical Basis}",
    eprint = "0808.1855",
    archivePrefix = "arXiv",
    primaryClass = "hep-ph",
    doi = "10.1088/0034-4885/72/2/026901",
    journal = "Rept. Prog. Phys.",
    volume = "72",
    pages = "026901",
    year = "2009"
}

@article{Kamali:2023lzq,
    author = "Kamali, Vahid and Motaharfar, Meysam and O. Ramos, Rudnei",
    title = "{Recent Developments in Warm Inflation}",
    eprint = "2302.02827",
    archivePrefix = "arXiv",
    primaryClass = "hep-ph",
    doi = "10.3390/universe9030124",
    journal = "Universe",
    volume = "9",
    number = "3",
    pages = "124",
    year = "2023"
}

@article{Yokoyama:1998ju,
    author = "Yokoyama, Junichi and Linde, Andrei D.",
    title = "{Is warm inflation possible?}",
    eprint = "hep-ph/9809409",
    archivePrefix = "arXiv",
    primaryClass = "hep-ph",
    reportNumber = "SU-ITP-98-52, YITP-98-49",
    doi = "10.1103/PhysRevD.60.083509",
    journal = "Phys. Rev. D",
    volume = "60",
    pages = "083509",
    year = "1999"
}

@article{Berghaus:2019whh,
    author = "Berghaus, Kim V. and Graham, Peter W. and Kaplan, David E.",
    title = "{Minimal Warm Inflation}",
    eprint = "1910.07525",
    archivePrefix = "arXiv",
    primaryClass = "hep-ph",
    doi = "10.1088/1475-7516/2020/03/034",
    journal = "JCAP",
    volume = "03",
    pages = "034",
    year = "2020",
    note = "[Erratum: JCAP 10, E02 (2023)]"
}

@article{McLerran:1990de,
    author = "McLerran, Larry D. and Mottola, Emil and Shaposhnikov, Mikhail E.",
    title = "{Sphalerons and Axion Dynamics in High Temperature {QCD}}",
    reportNumber = "TPI-MINN-90-15-T, LA-UR-90-800",
    doi = "10.1103/PhysRevD.43.2027",
    journal = "Phys. Rev. D",
    volume = "43",
    pages = "2027--2035",
    year = "1991"
}

@article{Laine:2021ego,
    author = "Laine, M. and Procacci, S.",
    title = "{Minimal warm inflation with complete medium response}",
    eprint = "2102.09913",
    archivePrefix = "arXiv",
    primaryClass = "hep-ph",
    doi = "10.1088/1475-7516/2021/06/031",
    journal = "JCAP",
    volume = "06",
    pages = "031",
    year = "2021"
}

@article{ORamos:2025uqs,
    author = "O. Ramos, Rudnei and Rodrigues, Gabriel S.",
    title = "{Viability of warm inflation with standard model interactions}",
    eprint = "2504.20943",
    archivePrefix = "arXiv",
    primaryClass = "hep-ph",
    doi = "10.1103/wn1m-19gt",
    journal = "Phys. Rev. D",
    volume = "111",
    number = "12",
    pages = "123527",
    year = "2025"
}

@article{Berghaus:2024zfg,
    author = "Berghaus, Kim V. and Forslund, Matthew and Guevarra, Mark Vincent",
    title = "{Warm inflation with a heavy QCD axion}",
    eprint = "2402.13535",
    archivePrefix = "arXiv",
    primaryClass = "hep-ph",
    doi = "10.1088/1475-7516/2024/10/103",
    journal = "JCAP",
    volume = "10",
    pages = "103",
    year = "2024"
}

@article{Drewes:2023khq,
    author = "Drewes, Marco and Zell, Sebastian",
    title = "{On sphaleron heating in the presence of fermions}",
    eprint = "2312.13739",
    archivePrefix = "arXiv",
    primaryClass = "hep-ph",
    doi = "10.1088/1475-7516/2024/06/038",
    journal = "JCAP",
    volume = "06",
    pages = "038",
    year = "2024"
}

@article{Gross:1980br,
    author = "Gross, David J. and Pisarski, Robert D. and Yaffe, Laurence G.",
    title = "{QCD and Instantons at Finite Temperature}",
    reportNumber = "PRINT-80-0538 (PRINCETON)",
    doi = "10.1103/RevModPhys.53.43",
    journal = "Rev. Mod. Phys.",
    volume = "53",
    pages = "43",
    year = "1981"
}

@article{Moore:2010jd,
    author = "Moore, Guy D. and Tassler, Marcus",
    title = "{The Sphaleron Rate in SU(N) Gauge Theory}",
    eprint = "1011.1167",
    archivePrefix = "arXiv",
    primaryClass = "hep-ph",
    doi = "10.1007/JHEP02(2011)105",
    journal = "JHEP",
    volume = "02",
    pages = "105",
    year = "2011"
}

@article{Khlebnikov:1988sr,
    author = "Khlebnikov, S.Yu. and Shaposhnikov, M.E.",
    title = "{The Statistical Theory of Anomalous Fermion Number Nonconservation}",
    doi = "10.1016/0550-3213(88)90133-2",
    journal = "Nucl. Phys. B",
    volume = "308",
    pages = "885--912",
    year = "1988"
}

@article{Moore:1996qs,
    author = "Moore, Guy D.",
    title = "{Motion of Chern-Simons number at high temperatures under a chemical potential}",
    eprint = "hep-ph/9603384",
    archivePrefix = "arXiv",
    primaryClass = "hep-ph",
    reportNumber = "PUPT-1606",
    doi = "10.1016/S0550-3213(96)00445-2",
    journal = "Nucl. Phys. B",
    volume = "480",
    pages = "657--688",
    year = "1996"
}

@book{Laine:2016hma,
    author = "Laine, Mikko and Vuorinen, Aleksi",
    title = "{Basics of Thermal Field Theory}",
    eprint = "1701.01554",
    archivePrefix = "arXiv",
    primaryClass = "hep-ph",
    doi = "10.1007/978-3-319-31933-9",
    publisher = "Springer",
    series = "Lect. Notes Phys.",
    volume = "925",
    year = "2016"
}

@article{Joyce:1997uy,
    author = "Joyce, M. and Shaposhnikov, Mikhail E.",
    title = "{Primordial magnetic fields, right-handed electrons, and the Abelian anomaly}",
    eprint = "astro-ph/9703005",
    archivePrefix = "arXiv",
    primaryClass = "astro-ph",
    reportNumber = "CERN-TH-97-31",
    doi = "10.1103/PhysRevLett.79.1193",
    journal = "Phys. Rev. Lett.",
    volume = "79",
    pages = "1193--1196",
    year = "1997"
}

@article{Akamatsu:2013pjd,
    author = "Akamatsu, Yukinao and Yamamoto, Naoki",
    title = "{Chiral Plasma Instabilities}",
    eprint = "1302.2125",
    archivePrefix = "arXiv",
    primaryClass = "nucl-th",
    reportNumber = "YITP-13-9",
    doi = "10.1103/PhysRevLett.111.052002",
    journal = "Phys. Rev. Lett.",
    volume = "111",
    pages = "052002",
    year = "2013"
}

@article{Anber:2009ua,
    author = "Anber, Mohamed M. and Sorbo, Lorenzo",
    title = "{Naturally inflating on steep potentials through electromagnetic dissipation}",
    eprint = "0908.4089",
    archivePrefix = "arXiv",
    primaryClass = "hep-th",
    doi = "10.1103/PhysRevD.81.043534",
    journal = "Phys. Rev. D",
    volume = "81",
    pages = "043534",
    year = "2010"
}

@article{Akamatsu:2015kau,
    author = "Akamatsu, Yukinao and Rothkopf, Alexander and Yamamoto, Naoki",
    title = "{Non-Abelian chiral instabilities at high temperature on the lattice}",
    eprint = "1512.02374",
    archivePrefix = "arXiv",
    primaryClass = "hep-ph",
    doi = "10.1007/JHEP03(2016)210",
    journal = "JHEP",
    volume = "03",
    pages = "210",
    year = "2016"
}

@article{Schlichting:2022fjc,
    author = "Schlichting, S{\"o}ren and Sharma, Sayantan",
    title = "{Chiral Instabilities and the Fate of Chirality Imbalance in Non-Abelian Plasmas}",
    eprint = "2211.11365",
    archivePrefix = "arXiv",
    primaryClass = "hep-ph",
    doi = "10.1103/PhysRevLett.131.102303",
    journal = "Phys. Rev. Lett.",
    volume = "131",
    number = "10",
    pages = "102303",
    year = "2023"
}

@article{Planck:2018vyg,
    author = "Aghanim, N. and others",
    collaboration = "Planck",
    title = "{Planck 2018 results. VI. Cosmological parameters}",
    eprint = "1807.06209",
    archivePrefix = "arXiv",
    primaryClass = "astro-ph.CO",
    doi = "10.1051/0004-6361/201833910",
    journal = "Astron. Astrophys.",
    volume = "641",
    pages = "A6",
    year = "2020",
    note = "[Erratum: Astron.Astrophys. 652, C4 (2021)]"
}

@article{Cohen:1987vi,
    author = "Cohen, Andrew G. and Kaplan, David B.",
    title = "{Thermodynamic Generation of the Baryon Asymmetry}",
    reportNumber = "HUTP-87/A061",
    doi = "10.1016/0370-2693(87)91369-4",
    journal = "Phys. Lett. B",
    volume = "199",
    pages = "251--258",
    year = "1987"
}

@article{Domcke:2020kcp,
    author = "Domcke, Valerie and Ema, Yohei and Mukaida, Kyohei and Yamada, Masaki",
    title = "{Spontaneous Baryogenesis from Axions with Generic Couplings}",
    eprint = "2006.03148",
    archivePrefix = "arXiv",
    primaryClass = "hep-ph",
    reportNumber = "DESY 20-100, DESY-20-100, CERN-TH-2020-088, TU-1102",
    doi = "10.1007/JHEP08(2020)096",
    journal = "JHEP",
    volume = "08",
    pages = "096",
    year = "2020"
}

@article{Brandenberger:2003kc,
    author = "Brandenberger, Robert H. and Yamaguchi, Masahide",
    title = "{Spontaneous baryogenesis in warm inflation}",
    eprint = "hep-ph/0301270",
    archivePrefix = "arXiv",
    primaryClass = "hep-ph",
    reportNumber = "BROWN-HET-1345",
    doi = "10.1103/PhysRevD.68.023505",
    journal = "Phys. Rev. D",
    volume = "68",
    pages = "023505",
    year = "2003"
}

@article{Basak:2021cgk,
    author = "Basak, Soumen and Bhattacharya, Sukannya and Gangopadhyay, Mayukh R. and Jaman, Nur and Rangarajan, Raghavan and Sami, M.",
    title = "{The paradigm of warm quintessential inflation and spontaneous baryogenesis}",
    eprint = "2110.00607",
    archivePrefix = "arXiv",
    primaryClass = "astro-ph.CO",
    doi = "10.1088/1475-7516/2022/03/063",
    journal = "JCAP",
    volume = "03",
    number = "03",
    pages = "063",
    year = "2022"
}

@article{Weinberg:1979sa,
    author = "Weinberg, Steven",
    title = "{Baryon and Lepton Nonconserving Processes}",
    reportNumber = "HUTP-79-A050",
    doi = "10.1103/PhysRevLett.43.1566",
    journal = "Phys. Rev. Lett.",
    volume = "43",
    pages = "1566--1570",
    year = "1979"
}

@article{Harvey:1990qw,
    author = "Harvey, Jeffrey A. and Turner, Michael S.",
    title = "{Cosmological Baryon and Lepton Number in the Presence of Electroweak Fermion Number Violation}",
    reportNumber = "FERMILAB-PUB-90-049-A, EFI-90-33",
    doi = "10.1103/PhysRevD.42.3344",
    journal = "Phys. Rev. D",
    volume = "42",
    pages = "3344--3349",
    year = "1990"
}

@article{Benson:1991nj,
    author = "Benson, Katherine M. and Bernstein, Jeremy and Dodelson, Scott",
    title = "{Phase structure and the effective potential at fixed charge}",
    reportNumber = "CERN-TH-6098-91, HUTP-91-A023",
    doi = "10.1103/PhysRevD.44.2480",
    journal = "Phys. Rev. D",
    volume = "44",
    pages = "2480--2497",
    year = "1991"
}

@article{Esteban:2024eli,
    author = "Esteban, Ivan and Gonzalez-Garcia, M.C. and Maltoni, Michele and Martinez-Soler, Ivan and Pinheiro, Jo{\~a}o Paulo and Schwetz, Thomas",
    title = "{NuFit-6.0: updated global analysis of three-flavor neutrino oscillations}",
    eprint = "2410.05380",
    archivePrefix = "arXiv",
    primaryClass = "hep-ph",
    reportNumber = "IFT-UAM/CSIC-24-140, YITP-SB-2024-24, IPPP/24/64",
    doi = "10.1007/JHEP12(2024)216",
    journal = "JHEP",
    volume = "12",
    pages = "216",
    year = "2024"
}

@article{Laine:2025rll,
    author = "Laine, M. and Procacci, S. and Rogelj, A.",
    title = "{Evolution of coupled scalar perturbations through smooth reheating. Part II. Thermal fluctuation regime}",
    eprint = "2507.12849",
    archivePrefix = "arXiv",
    primaryClass = "hep-ph",
    doi = "10.1088/1475-7516/2025/12/058",
    journal = "JCAP",
    volume = "12",
    pages = "058",
    year = "2025"
}

@article{Bodeker:1998hm,
    author = "B{\"o}deker, Dietrich",
    title = "{On the effective dynamics of soft non-Abelian gauge fields at finite temperature}",
    eprint = "hep-ph/9801430",
    archivePrefix = "arXiv",
    primaryClass = "hep-ph",
    journal = "Phys. Lett. B",
    volume = "426",
    pages = "351--360",
    year = "1998"
}

@article{Buttazzo:2013uya,
    author = "Buttazzo, Dario and Degrassi, Giuseppe and Giardino, Pier Paolo and Giudice, Gian F. and Sala, Filippo and Salvio, Alberto and Strumia, Alessandro",
    title = "{Investigating the near-criticality of the Higgs boson}",
    eprint = "1307.3536",
    archivePrefix = "arXiv",
    primaryClass = "hep-ph",
    journal = "JHEP",
    volume = "12",
    pages = "089",
    year = "2013"
}

@article{Espinosa:2007qp,
    author = "Espinosa, J. R. and Giudice, G. F. and Riotto, A.",
    title = "{Cosmological implications of the Higgs mass measurement}",
    eprint = "0710.2484",
    archivePrefix = "arXiv",
    primaryClass = "hep-ph",
    journal = "JCAP",
    volume = "05",
    pages = "002",
    year = "2008"
}

@article{Berghaus:2025dqi,
    author = "Berghaus, Kim V. and Drewes, Marco and Zell, Sebastian",
    title = "{Warm Inflation with the Standard Model}",
    eprint = "2503.18829",
    archivePrefix = "arXiv",
    primaryClass = "hep-ph",
    doi = "10.1103/9nn9-bsm9",
    journal = "Phys. Rev. Lett.",
    volume = "135",
    number = "17",
    pages = "171002",
    year = "2025"
}

@article{Kubo:1957mj,
    author = "Kubo, Ryogo",
    title = "{Statistical mechanical theory of irreversible processes. 1. General theory and simple applications in magnetic and conduction problems}",
    doi = "10.1143/JPSJ.12.570",
    journal = "J. Phys. Soc. Jap.",
    volume = "12",
    pages = "570--586",
    year = "1957"
}

@article{Bodeker:1999gx,
    author = "B{\"o}deker, D. and Moore, Guy D. and Rummukainen, K.",
    title = "{Chern-Simons number diffusion and hard thermal loops on the lattice}",
    eprint = "hep-ph/9907545",
    archivePrefix = "arXiv",
    doi = "10.1103/PhysRevD.61.056003",
    journal = "Phys. Rev. D",
    volume = "61",
    pages = "056003",
    year = "2000"
}

@article{Arnold:1996dy,
    author = "Arnold, Peter Brockway and Son, Dam and Yaffe, Laurence G.",
    title = "{The Hot baryon violation rate is $O(\alpha_w^5 T^4)$}",
    eprint = "hep-ph/9609481",
    archivePrefix = "arXiv",
    doi = "10.1103/PhysRevD.55.6264",
    journal = "Phys. Rev. D",
    volume = "55",
    pages = "6264--6273",
    year = "1997"
}

@article{Arnold:2002zm,
    author = "Arnold, Peter Brockway and Moore, Guy D. and Yaffe, Laurence G.",
    title = "{Effective kinetic theory for high temperature gauge theories}",
    eprint = "hep-ph/0209353",
    archivePrefix = "arXiv",
    doi = "10.1088/1126-6708/2003/01/030",
    journal = "JHEP",
    volume = "01",
    pages = "030",
    year = "2003"
}

@article{tHooft:1976rip,
    author = "'t Hooft, Gerard",
    title = "{Symmetry Breaking Through Bell-Jackiw Anomalies}",
    doi = "10.1103/PhysRevLett.37.8",
    journal = "Phys. Rev. Lett.",
    volume = "37",
    pages = "8--11",
    year = "1976"
}

@article{Kuzmin:1985mm,
    author = "Kuzmin, V. A. and Rubakov, V. A. and Shaposhnikov, M. E.",
    title = "{On the Anomalous Electroweak Baryon Number Nonconservation in the Early Universe}",
    doi = "10.1016/0370-2693(85)91028-7",
    journal = "Phys. Lett. B",
    volume = "155",
    pages = "36",
    year = "1985"
}

@article{Rubakov:1996vz,
    author = "Rubakov, V. A. and Shaposhnikov, M. E.",
    title = "{Electroweak baryon number nonconservation in the early universe and in high-energy collisions}",
    eprint = "hep-ph/9603208",
    archivePrefix = "arXiv",
    doi = "10.1070/PU1996v039n05ABEH000145",
    journal = "Usp. Fiz. Nauk",
    volume = "166",
    pages = "493--537",
    year = "1996",
    note = "[Phys. Usp. 39, 461 (1996)]"
}

@article{Kuzmin:1992up,
    author = "Kuzmin, V. A. and Shaposhnikov, M. E. and Tkachev, I. I.",
    title = "{Strong CP violation, electroweak baryogenesis, and axionic dark matter}",
    doi = "10.1103/PhysRevD.45.466",
    journal = "Phys. Rev. D",
    volume = "45",
    pages = "466--475",
    year = "1992"
}

@article{Berghaus:2020ekh,
    author = "Berghaus, Kim V. and Graham, Peter W. and Kaplan, David E. and Moore, Guy D. and Rajendran, Surjeet",
    title = "{Dark energy radiation}",
    eprint = "2012.10549",
    archivePrefix = "arXiv",
    primaryClass = "hep-ph",
    doi = "10.1103/PhysRevD.104.083520",
    journal = "Phys. Rev. D",
    volume = "104",
    pages = "083520",
    year = "2021"
}

@article{Arnold:1998cy,
    author = "Arnold, Peter Brockway and Son, Dam T. and Yaffe, Laurence G.",
    title = "{Effective dynamics of hot, soft nonAbelian gauge fields. Color conductivity and log(1/alpha) effects}",
    eprint = "hep-ph/9810216",
    archivePrefix = "arXiv",
    doi = "10.1103/PhysRevD.59.105020",
    journal = "Phys. Rev. D",
    volume = "59",
    pages = "105020",
    year = "1999"
}

@article{Broadberry:2025ggb,
    author = "Broadberry, Edward and Hook, Anson and Mondal, Sagnik",
    title = "{Warm inflation with pseudo-scalar couplings}",
    eprint = "2505.07943",
    archivePrefix = "arXiv",
    primaryClass = "hep-ph",
    doi = "10.1007/JHEP08(2026)072",
    journal = "JHEP",
    volume = "08",
    pages = "072",
    year = "2026"
}

@article{Dejrah:2026ltr,
    author = "Dejrah, Rafid H. and Ghoshal, Anish",
    title = "{Gauge Backreaction in Standard Model Warm Inflation}",
    eprint = "2609.35756",
    archivePrefix = "arXiv",
    primaryClass = "astro-ph.CO",
    month = "9",
    year = "2026"
}

@article{Arnold:2003zc,
    author = "Arnold, Peter Brockway and Moore, Guy D. and Yaffe, Laurence G.",
    title = "{Transport coefficients in high temperature gauge theories. 2. Beyond leading log}",
    eprint = "hep-ph/0302165",
    archivePrefix = "arXiv",
    doi = "10.1088/1126-6708/2003/05/051",
    journal = "JHEP",
    volume = "05",
    pages = "051",
    year = "2003"
}

@article{York:2008rr,
    author = "York, Mark Abraao and Moore, Guy D.",
    title = "{Second order hydrodynamic coefficients from kinetic theory}",
    eprint = "0811.0729",
    archivePrefix = "arXiv",
    primaryClass = "hep-ph",
    doi = "10.1103/PhysRevD.79.054011",
    journal = "Phys. Rev. D",
    volume = "79",
    pages = "054011",
    year = "2009"
}

@article{Fujita:2025zoa,
    author = "Fujita, Tomohiro and Mukaida, Kyohei and Tsuji, Tenta",
    title = "{Reheating after axion inflation}",
    eprint = "2503.01228",
    archivePrefix = "arXiv",
    doi = "10.1088/1475-7516/2025/07/002",
    journal = "JCAP",
    volume = "07",
    pages = "002",
    year = "2025"
}

@article{Hiller:2024zjp,
    author = {Hiller, Gudrun and H{\"o}hne, Tim and Litim, Daniel F. and Steudtner, Tom},
    title = "{Vacuum stability in the Standard Model and beyond}",
    eprint = "2401.08811",
    archivePrefix = "arXiv",
    primaryClass = "hep-ph",
    doi = "10.1103/PhysRevD.110.115017",
    journal = "Phys. Rev. D",
    volume = "110",
    number = "11",
    pages = "115017",
    year = "2024"
}

@article{Kuckenberg:2026oax,
    author = "Kuckenberg, Konstantin and Mojahed, Martin A. and Schmitz, Kai and Watanabe, Hidenaga",
    title = "{Axion-driven spontaneous leptogenesis, precisely}",
    eprint = "2608.05279",
    archivePrefix = "arXiv",
    primaryClass = "hep-ph",
    reportNumber = "MS-TP-26-19, KEK-TH-2858",
    month = "8",
    year = "2026"
}

@article{Garbrecht:2014kda,
    author = {Garbrecht, Bj{\"o}rn and Schwaller, Pedro},
    title = "{Spectator Effects during Leptogenesis in the Strong Washout Regime}",
    eprint = "1404.2915",
    archivePrefix = "arXiv",
    primaryClass = "hep-ph",
    reportNumber = "TUM-HEP-938-14, CERN-PH-TH-2014-059",
    doi = "10.1088/1475-7516/2014/10/012",
    journal = "JCAP",
    volume = "10",
    pages = "012",
    year = "2014"
}

@article{Heller:1997nqa,
    author = "Heller, Urs M. and Karsch, Frithjof and Rank, J.",
    title = "{The gluon propagator at high temperature: Screening, improvement and nonzero momenta}",
    eprint = "hep-lat/9710033",
    archivePrefix = "arXiv",
    doi = "10.1103/PhysRevD.57.1438",
    journal = "Phys. Rev. D",
    volume = "57",
    pages = "1438--1448",
    year = "1998"
}

@article{Nakamura:2003pu,
    author = "Nakamura, A. and Saito, T. and Sakai, S.",
    title = "{Lattice calculation of gluon screening masses}",
    eprint = "hep-lat/0311024",
    archivePrefix = "arXiv",
    doi = "10.1103/PhysRevD.69.014506",
    journal = "Phys. Rev. D",
    volume = "69",
    pages = "014506",
    year = "2004"
}

@article{Hook:2019vcn,
    author = "Hook, Anson and Huang, Junwu and Racco, Davide",
    title = "{Minimal signatures of the Standard Model in non-Gaussianities}",
    eprint = "1908.00019",
    archivePrefix = "arXiv",
    primaryClass = "hep-ph",
    doi = "10.1103/PhysRevD.101.023519",
    journal = "Phys. Rev. D",
    volume = "101",
    number = "2",
    pages = "023519",
    year = "2020"
}

@article{Berghaus:2026jkw,
    author = "Berghaus, Kim V. and Florio, Adrien and Laine, M. and Sattler, Franz R.",
    title = "{Classical equipartition dynamics between axions and non-Abelian gauge fields}",
    eprint = "2601.09784",
    archivePrefix = "arXiv",
    primaryClass = "hep-ph",
    reportNumber = "CALT-TH/2026-002",
    doi = "10.1088/1475-7516/2026/06/020",
    journal = "JCAP",
    volume = "06",
    pages = "020",
    year = "2026"
}

@article{Ghoshal:2026qio,
    author = "Ghoshal, Anish and Strumia, Alessandro",
    title = "{Challenges for minimal warm inflation in the Standard Model and beyond}",
    eprint = "2611.09893",
    archivePrefix = "arXiv",
    primaryClass = "hep-ph",
    month = "11",
    year = "2026"
}

@article{Peccei:1977hh,
    author = "Peccei, R. D. and Quinn, Helen R.",
    title = "{CP Conservation in the Presence of Instantons}",
    reportNumber = "ITP-568-STANFORD",
    doi = "10.1103/PhysRevLett.38.1440",
    journal = "Phys. Rev. Lett.",
    volume = "38",
    pages = "1440--1443",
    year = "1977"
}

@article{Cohen:1988kt,
    author = "Cohen, Andrew G. and Kaplan, David B.",
    title = "{SPONTANEOUS BARYOGENESIS}",
    reportNumber = "HUTP-88/A016",
    doi = "10.1016/0550-3213(88)90134-4",
    journal = "Nucl. Phys. B",
    volume = "308",
    pages = "913--928",
    year = "1988"
}

@article{Foot:1988aq,
    author = "Foot, Robert and Lew, H. and He, X. G. and Joshi, Girish C.",
    title = "{Seesaw Neutrino Masses Induced by a Triplet of Leptons}",
    reportNumber = "UM-P-88/89, OZ-P-88/7",
    doi = "10.1007/BF01415558",
    journal = "Z. Phys. C",
    volume = "44",
    pages = "441",
    year = "1989"
}

@article{Ma:1998dn,
    author = "Ma, Ernest",
    title = "{Pathways to naturally small neutrino masses}",
    eprint = "hep-ph/9805219",
    archivePrefix = "arXiv",
    reportNumber = "UCRHEP-T-222",
    doi = "10.1103/PhysRevLett.81.1171",
    journal = "Phys. Rev. Lett.",
    volume = "81",
    pages = "1171--1174",
    year = "1998"
}

@article{Schechter:1980gr,
    author = "Schechter, J. and Valle, J. W. F.",
    title = "{Neutrino Masses in SU(2) x U(1) Theories}",
    reportNumber = "SU-4217-167, COO-3533-167",
    doi = "10.1103/PhysRevD.22.2227",
    journal = "Phys. Rev. D",
    volume = "22",
    pages = "2227",
    year = "1980"
}

@article{Magg:1980ut,
    author = "Magg, M. and Wetterich, C.",
    title = "{Neutrino Mass Problem and Gauge Hierarchy}",
    reportNumber = "CERN-TH-2829",
    doi = "10.1016/0370-2693(80)90825-4",
    journal = "Phys. Lett. B",
    volume = "94",
    pages = "61--64",
    year = "1980"
}

@article{Lazarides:1980nt,
    author = "Lazarides, George and Shafi, Q. and Wetterich, C.",
    title = "{Proton Lifetime and Fermion Masses in an SO(10) Model}",
    reportNumber = "FREIBURG-THEP-80-2",
    doi = "10.1016/0550-3213(81)90354-0",
    journal = "Nucl. Phys. B",
    volume = "181",
    pages = "287--300",
    year = "1981"
}

@article{Mohapatra:1980yp,
    author = "Mohapatra, Rabindra N. and Senjanovic, Goran",
    title = "{Neutrino Masses and Mixings in Gauge Models with Spontaneous Parity Violation}",
    reportNumber = "FERMILAB-PUB-80-061-THY, FERMILAB-PUB-80-061-T",
    doi = "10.1103/PhysRevD.23.165",
    journal = "Phys. Rev. D",
    volume = "23",
    pages = "165",
    year = "1981"
}

@article{Cheng:1980qt,
    author = "Cheng, T. P. and Li, Ling-Fong",
    title = "{Neutrino Masses, Mixings and Oscillations in SU(2) x U(1) Models of Electroweak Interactions}",
    reportNumber = "PRINT-80-0511 (CARNEGIE-MELLON), COO-3066-152",
    doi = "10.1103/PhysRevD.22.2860",
    journal = "Phys. Rev. D",
    volume = "22",
    pages = "2860",
    year = "1980"
}

@article{Minkowski:1977sc,
    author = "Minkowski, Peter",
    title = "{$\mu \to e\gamma$ at a Rate of One Out of $10^{9}$ Muon Decays?}",
    reportNumber = "Print-77-0182 (BERN)",
    doi = "10.1016/0370-2693(77)90435-X",
    journal = "Phys. Lett. B",
    volume = "67",
    pages = "421--428",
    year = "1977"
}

@article{Yanagida:1979as,
    author = "Yanagida, Tsutomu",
    editor = "Sawada, Osamu and Sugamoto, Akio",
    title = "{Horizontal gauge symmetry and masses of neutrinos}",
    reportNumber = "KEK-79-18-95",
    journal = "Conf. Proc. C",
    volume = "7902131",
    pages = "95--99",
    year = "1979"
}

@article{Yanagida:1980xy,
    author = "Yanagida, Tsutomu",
    title = "{Horizontal Symmetry and Masses of Neutrinos}",
    reportNumber = "TU-80-208",
    doi = "10.1143/PTP.64.1103",
    journal = "Prog. Theor. Phys.",
    volume = "64",
    pages = "1103",
    year = "1980"
}

@article{Gell-Mann:1979vob,
    author = "Gell-Mann, Murray and Ramond, Pierre and Slansky, Richard",
    title = "{Complex Spinors and Unified Theories}",
    eprint = "1306.4669",
    archivePrefix = "arXiv",
    primaryClass = "hep-th",
    reportNumber = "PRINT-80-0576",
    journal = "Conf. Proc. C",
    volume = "790927",
    pages = "315--321",
    year = "1979"
}

@article{Mohapatra:1979ia,
    author = "Mohapatra, Rabindra N. and Senjanovic, Goran",
    title = "{Neutrino Mass and Spontaneous Parity Nonconservation}",
    reportNumber = "MDDP-TR-80-060, MDDP-PP-80-105, CCNY-HEP-79-10",
    doi = "10.1103/PhysRevLett.44.912",
    journal = "Phys. Rev. Lett.",
    volume = "44",
    pages = "912",
    year = "1980"
}

@article{Bastero-Gil:2011clw,
    author = "Bastero-Gil, Mar and Berera, Arjun and Ramos, Rudnei O. and Rosa, Joao G.",
    title = "{Warm baryogenesis}",
    eprint = "1110.3971",
    archivePrefix = "arXiv",
    primaryClass = "hep-ph",
    reportNumber = "EDINBURGH-2011-29, UG-FT-291-11, CFPE-161-11",
    doi = "10.1016/j.physletb.2012.05.032",
    journal = "Phys. Lett. B",
    volume = "712",
    pages = "425--429",
    year = "2012"
}

@article{Benso:2026csa,
    author = {Benso, Cristina and B{\"o}deker, Dietrich and Kamada, Kohei and Mukaida, Kyohei and Sagunski, Laura and Schicho, Philipp and Schmitz, Kai},
    title = "{Baryon number freeze-out in the Standard Model, precisely}",
    eprint = "2609.07467",
    archivePrefix = "arXiv",
    primaryClass = "hep-ph",
    reportNumber = "MS-TP-26-25, KEK-TH-2869",
    month = "9",
    year = "2026"
}

@article{Laine:1999wv,
    author = "Laine, M. and Shaposhnikov, Mikhail E.",
    title = "{A Remark on sphaleron erasure of baryon asymmetry}",
    eprint = "hep-ph/9911473",
    archivePrefix = "arXiv",
    reportNumber = "CERN-TH-99-353, UNIL-IPT-99-6",
    doi = "10.1103/PhysRevD.61.117302",
    journal = "Phys. Rev. D",
    volume = "61",
    pages = "117302",
    year = "2000"
}

@article{Sereewat:2026kbc,
    author = "Sereewat, Phongsakorn",
    title = "{Gravitational Production of Non-minimally Coupled Vector Dark Matter in Warm Inflation}",
    eprint = "2611.10090",
    archivePrefix = "arXiv",
    primaryClass = "astro-ph.CO",
    month = "11",
    year = "2026"
}

@article{Biondini:2026wfy,
    author = "Biondini, Simone and Laine, M.",
    title = "{Superhorizon enhancement of the hydrodynamic GW tail in Standard Model embedded warm inflation}",
    eprint = "2610.19500",
    archivePrefix = "arXiv",
    primaryClass = "hep-ph",
    month = "10",
    year = "2026"
}

@phdthesis{Rogelj:2026kgr,
    author = "Rogelj, Alica Ela",
    title = "{Scalar perturbations in warm inflation and during reheating}",
    eprint = "2610.11880",
    archivePrefix = "arXiv",
    primaryClass = "hep-ph",
    doi = "10.48620/101166",
    school = "Bern U.",
    year = "2026"
}

@article{Starobinsky:1980te,
    author = "Starobinsky, Alexei A.",
    editor = "Khalatnikov, I. M. and Mineev, V. P.",
    title = "{A New Type of Isotropic Cosmological Models Without Singularity}",
    doi = "10.1016/0370-2693(80)90670-X",
    journal = "Phys. Lett. B",
    volume = "91",
    pages = "99--102",
    year = "1980"
}

@article{Guth:1980zm,
    author = "Guth, Alan H.",
    editor = "Fang, Li-Zhi and Ruffini, R.",
    title = "{The Inflationary Universe: A Possible Solution to the Horizon and Flatness Problems}",
    reportNumber = "SLAC-PUB-2576",
    doi = "10.1103/PhysRevD.23.347",
    journal = "Phys. Rev. D",
    volume = "23",
    pages = "347--356",
    year = "1981"
}

@article{Linde:1981mu,
    author = "Linde, Andrei D.",
    editor = "Fang, Li-Zhi and Ruffini, R.",
    title = "{A New Inflationary Universe Scenario: A Possible Solution of the Horizon, Flatness, Homogeneity, Isotropy and Primordial Monopole Problems}",
    reportNumber = "LEBEDEV-81-229",
    doi = "10.1016/0370-2693(82)91219-9",
    journal = "Phys. Lett. B",
    volume = "108",
    pages = "389--393",
    year = "1982"
}

@article{Mukhanov:1981xt,
    author = "Mukhanov, Viatcheslav F. and Chibisov, G. V.",
    title = "{Quantum Fluctuations and a Nonsingular Universe}",
    journal = "JETP Lett.",
    volume = "33",
    pages = "532--535",
    year = "1981"
}

@article{Bardeen:1983qw,
    author = "Bardeen, James M. and Steinhardt, Paul J. and Turner, Michael S.",
    title = "{Spontaneous Creation of Almost Scale - Free Density Perturbations in an Inflationary Universe}",
    reportNumber = "UPR-0202T, EFI-83-13-CHICAGO",
    doi = "10.1103/PhysRevD.28.679",
    journal = "Phys. Rev. D",
    volume = "28",
    pages = "679",
    year = "1983"
}

@article{Albrecht:1982wi,
    author = "Albrecht, Andreas and Steinhardt, Paul J.",
    editor = "Fang, Li-Zhi and Ruffini, R.",
    title = "{Cosmology for Grand Unified Theories with Radiatively Induced Symmetry Breaking}",
    reportNumber = "UPR-0185T",
    doi = "10.1103/PhysRevLett.48.1220",
    journal = "Phys. Rev. Lett.",
    volume = "48",
    pages = "1220--1223",
    year = "1982"
}

@article{Bastero-Gil:2016qru,
    author = "Bastero-Gil, Mar and Berera, Arjun and Ramos, Rudnei O. and Rosa, Joao G.",
    title = "{Warm Little Inflaton}",
    eprint = "1604.08838",
    archivePrefix = "arXiv",
    primaryClass = "hep-ph",
    doi = "10.1103/PhysRevLett.117.151301",
    journal = "Phys. Rev. Lett.",
    volume = "117",
    number = "15",
    pages = "151301",
    year = "2016"
}

@article{Chen:2025try,
    author = "Chen, Quan and Jiang, Siyu and Qiu, Dayun and Chen, Peilin and Huang, Fa Peng",
    title = "{Freeze-in gravitational waves and dark matter in warm inflation}",
    eprint = "2507.13916",
    archivePrefix = "arXiv",
    primaryClass = "hep-ph",
    doi = "10.1088/1475-7516/2026/03/051",
    journal = "JCAP",
    volume = "03",
    pages = "051",
    year = "2026"
}

@article{Bhattacharya:2025guc,
    author = "Bhattacharya, Sukannya and Fasiello, Matteo and Papageorgiou, Alexandros and Dimastrogiovanni, Ema",
    title = "{On the prospects of thermalization of axion-SU(2) inflation}",
    eprint = "2506.11853",
    archivePrefix = "arXiv",
    primaryClass = "astro-ph.CO",
    doi = "10.1088/1475-7516/2025/10/080",
    journal = "JCAP",
    volume = "10",
    pages = "080",
    year = "2025"
}

@article{Ferreira:2017lnd,
    author = "Ferreira, Ricardo Z. and Notari, Alessio",
    title = "{Thermalized Axion Inflation}",
    eprint = "1706.00373",
    archivePrefix = "arXiv",
    primaryClass = "astro-ph.CO",
    doi = "10.1088/1475-7516/2017/09/007",
    journal = "JCAP",
    volume = "09",
    pages = "007",
    year = "2017"
}

@article{DeRocco:2021rzv,
    author = "DeRocco, William and Graham, Peter W. and Kalia, Saarik",
    title = "{Warming up cold inflation}",
    eprint = "2107.07517",
    archivePrefix = "arXiv",
    primaryClass = "hep-ph",
    reportNumber = "FERMILAB-PUB-21-0919-SQMS-V",
    doi = "10.1088/1475-7516/2021/11/011",
    journal = "JCAP",
    volume = "11",
    pages = "011",
    year = "2021"
}

@article{Levy:2020zfo,
    author = "Levy, Miguel and Rosa, Jo{\~a}o G. and Ventura, Luis B.",
    title = "{Warm inflation, neutrinos and dark matter: a minimal extension of the Standard Model}",
    eprint = "2012.03988",
    archivePrefix = "arXiv",
    primaryClass = "hep-ph",
    doi = "10.1007/JHEP12(2021)176",
    journal = "JHEP",
    volume = "12",
    pages = "176",
    year = "2021"
}

@article{Chikashige:1980ui,
    author = "Chikashige, Y. and Mohapatra, Rabindra N. and Peccei, R. D.",
    title = "{Are There Real Goldstone Bosons Associated with Broken Lepton Number?}",
    reportNumber = "MPI-PAE-PTH-36-80",
    doi = "10.1016/0370-2693(81)90011-3",
    journal = "Phys. Lett. B",
    volume = "98",
    pages = "265--268",
    year = "1981"
}

@article{Bajc:1997ky,
    author = "Bajc, Borut and Riotto, Antonio and Senjanovic, Goran",
    title = "{Large lepton number of the universe and the fate of topological defects}",
    eprint = "hep-ph/9710415",
    archivePrefix = "arXiv",
    reportNumber = "DTP-97-92, IJS-TP-97-16, OUTP-97-49-P",
    doi = "10.1103/PhysRevLett.81.1355",
    journal = "Phys. Rev. Lett.",
    volume = "81",
    pages = "1355--1358",
    year = "1998"
}

@article{Linde:1976kh,
    author = "Linde, Andrei D.",
    title = "{High Density and High Temperature Symmetry Behavior in Gauge Theories}",
    reportNumber = "IC/76/27",
    doi = "10.1103/PhysRevD.14.3345",
    journal = "Phys. Rev. D",
    volume = "14",
    pages = "3345",
    year = "1976"
}

@article{Alekhin:2012py,
    author = "Alekhin, S. and Djouadi, A. and Moch, S.",
    title = "{The top quark and Higgs boson masses and the stability of the electroweak vacuum}",
    eprint = "1207.0980",
    archivePrefix = "arXiv",
    primaryClass = "hep-ph",
    reportNumber = "CERN-PH-TH-2012-186, DESY-12-113, LPT-ORSAY-12-64, LPN-12-073, SFB-CPP-12-41",
    doi = "10.1016/j.physletb.2012.08.024",
    journal = "Phys. Lett. B",
    volume = "716",
    pages = "214--219",
    year = "2012"
}

@article{Hoang:2020iah,
    author = "Hoang, Andr{\'e} H.",
    title = "{What is the Top Quark Mass?}",
    eprint = "2004.12915",
    archivePrefix = "arXiv",
    primaryClass = "hep-ph",
    reportNumber = "UWThPh-2020-1",
    doi = "10.1146/annurev-nucl-101918-023530",
    journal = "Ann. Rev. Nucl. Part. Sci.",
    volume = "70",
    pages = "225--255",
    year = "2020"
}

@article{Zell:2024cyz,
    author = "Zell, Sebastian",
    title = "{No warm inflation from a vanilla axion}",
    eprint = "2408.07746",
    archivePrefix = "arXiv",
    primaryClass = "hep-ph",
    doi = "10.1103/kp15-3s1c",
    journal = "Phys. Rev. D",
    volume = "112",
    number = "8",
    pages = "L081307",
    year = "2025"
}

\end{document}